\documentclass[USenglish]{article}	
\usepackage[utf8]{inputenc}
\usepackage{amsmath, amssymb, amsfonts}
\usepackage{geometry}
\usepackage{booktabs}
\usepackage{hyperref}
\usepackage{cite}
\usepackage{authblk}
\usepackage{adjustbox}
\usepackage{graphicx}
\usepackage[title]{appendix}
\usepackage{longtable}
\usepackage{multirow}
\usepackage{xcolor}
\usepackage{natbib}
\usepackage{rotating}

\title{\textbf{Simultaneous Estimation of Ballpark Effects and Team Defense Using Total Bases Residuals}}

\author[1]{Jhe-Jia Wu}
\author[2, 3]{Tian-Li Yan}
\author[1, 3]{Ting-Li Chen$^*$}

\affil[1]{Data Science Degree Program, National Taiwan University and Academia Sinica, Taipei, Taiwan}
\affil[2]{Department of Mathematics, University of Houston, Houston, USA}
\affil[3]{Institute of Statistical Science, Academia Sinica, Taipei, Taiwan}

\date{\today}

\begin{document}

\maketitle

\begin{abstract}
Park effect estimates and defensive metrics in baseball are typically constructed independently, even though both are shaped by the same batted-ball outcomes. Without jointly controlling for these factors, each may absorb variation that properly belongs to the other. We propose a framework based on Total Bases Residuals (TBR), defined as the departure of each batted ball's outcome from an expected baseline conditional on exit velocity and launch angle. Modeling these residuals with ballpark and defensive-team indicators in a weighted least-squares regression estimates both effects simultaneously using Statcast data from 2015 to 2024. When our ballpark estimates differ from official MLB park factors, home--away patterns for both teams and opponents are more consistent with our estimates, suggesting official metrics may partly reflect defensive or personnel effects. For team defense, our estimates align more closely with Outs Above Average than Defensive Runs Above Average. The league-wide intercept declines from 2015 to 2022, consistent with data-driven positioning and changes in ball construction, before reversing after the 2023 infield shift restrictions. We also introduce 100-based standardized indices with 95\% confidence intervals; for single-season park effects, the average half-width is about 20 points, indicating that sampling variability may explain part of the year-to-year instability.\footnote{Code is available at: \url{https://github.com/qqaazz800624/sports-science.git}}

\vspace{1em} % 增加一點與上方摘要本文的垂直距離
\noindent \textbf{Keywords:} Statcast, park factor, team defense, batted-ball outcomes, baseball analytics
\end{abstract}
	
\section{Introduction}
In the 2021 season, official MLB park factors rated Target Field as slightly favorable to hitters.  Yet direct comparisons of the same teams across locations tell a different story. Both the Minnesota Twins and their opponents show lower offensive production in Target Field than in other ballparks (see Appendix Table~\ref{tab:min_statcast_2021}). A ballpark that official statistics labeled as hitter-friendly appears, in these comparisons, to be suppressing offensive production. However, if we look at all games without separating teams, the pattern can be very different, and Target Field may appear to have strong offensive output. One possible reason is that Minnesota had one of the weaker pitching staffs that season. The discrepancy arises because observed offensive statistics mix together the ballpark environment, defensive performance, and team composition. Without separating these components, park effect estimates can point in the wrong direction.

The main difficulty is that the outcome of a batted ball depends on several factors at the same time. It depends on the quality of contact, the ballpark, and the defense on the field. These factors all act on the same outcome, so they are mixed together in the data. Because of this, it is hard to tell what causes the differences we observe. A park may look hitter-friendly because of weak pitching or poor defense, while a team may appear to have strong defense partly because of its home park. Without separating these effects, both park factors and defensive measures can be biased.

%\citep{Acharya2008, StatcastPF}.

Early methods often used simple home--road comparisons, as introduced in early sabermetric work by Bill James and later discussed in the sabermetric literature \citep{tango2007book}. However, these approaches rely on strong assumptions and can lead to bias. \citet{Acharya2008} showed that such methods can be seriously distorted when these assumptions fail. To address this problem, later work used regression models to estimate team strength and park effects at the same time \citep{Acharya2008,baumer2015openwar}. More recent studies further showed that even these models can still reflect team-specific characteristics or player composition \citep{osborne2026personnel}. However, these approaches focus mainly on estimating park effects.

Measuring defensive performance presents a closely related problem. Early statistics, such as errors and fielding percentage, capture only part of defensive ability. Academic work has developed spatial and temporal models to evaluate fielding performance more carefully \citep{james2012estimating}. These ideas form the basis of widely used metrics such as Ultimate Zone Rating (UZR), Defensive Runs Saved (DRS), Outs Above Average (OAA), and Defensive Runs Above Average (Def) (see, e.g., FanGraphs; MLB Statcast). However, these measures are typically developed separately from park-effect estimation, even though both defense and the ballpark affect the same batted-ball outcomes. As a result, it remains difficult to separate their contributions clearly.

In this study, we propose a simple framework to separate these effects. We use Statcast data and focus on exit velocity (EV) and launch angle (LA) to control for the quality of contact. For batted balls with similar contact, we define an expected level of offensive outcome and measure the difference between the observed and expected values. These differences, which we call total bases residuals (TBR), capture the remaining variation in the data.

We then model these residuals using a regression framework with ballpark and defensive team effects. This allows us to estimate ballpark effects and team defense at the same time. Aggregating observations with identical design rows improves computational efficiency without changing the coefficient estimates. The resulting estimates are simple to interpret and can be compared across teams, ballparks, and seasons. The regression framework also provides standard errors and confidence intervals, allowing these comparisons to be interpreted relative to sampling variability. Our results show that these estimates can differ substantially from official metrics and are often more consistent with observed home–away patterns. The uncertainty analysis further indicates that sampling variability may account for part of the apparent year-to-year instability in single-season park factors.

\section{Methodology}\label{sec:method}

\subsection{Outcome Residuals Conditional on Exit Velocity and Launch Angle}\label{sec:TBR}
In this study, we define the outcome of a batted ball using total bases rather than an out-based measure. Out-based approaches, such as those underlying OAA, focus on the probability that a batted ball is converted into an out. While such measures capture defensive success in terms of outs converted, they do not distinguish between different offensive outcomes once a ball is not converted into an out. 

In contrast, total bases capture the magnitude of offensive outcomes and provide a more informative measure of how batted balls contribute to run production. Because it is well known that measures based on extra-base hits are more strongly associated with run production than those based only on hit frequency, total bases are more appropriate when the goal is to assess how ballpark environment and defensive performance affect run production.

This distinction is also important relative to conventional measures such as slugging percentage. Because slugging percentage is defined per at-bat, it is affected by the frequency of strikeouts and therefore mixes the probability of putting the ball in play with the outcome of contact. Our analysis instead conditions on a batted-ball event and focuses on the number of bases produced after contact, which is more directly aligned with the ballpark and defensive effects considered here.

To separate the effects of ballpark and defense from batted-ball outcomes, we control for the quality of contact using exit velocity (EV) and launch angle (LA). Batted balls with similar EV and LA tend to produce similar outcomes on average, so grouping observations with similar contact
characteristics provides a natural way to remove the primary effect of contact quality. The remaining variation can then be used to study the combined influence of ballpark and defensive factors.

Having defined the outcome, we now turn to how we control for the quality of contact. We implement this idea by partitioning the EV–LA space into a fine grid and constructing an empirical distribution of outcomes within each grid cell. This approach of discretizing the physical properties of batted balls, formally known as measurement space partitioning, has been shown to significantly improve the modeling and prediction of batted-ball distributions \citep{healey2021measurement}. Furthermore, as \citet{healey2021measurement} noted, outcomes within these partitioned subregions are still subject to contextual factors such as outfield geometry and the quality of defenders, which perfectly motivates our residual-based decomposition. In our implementation, exit velocity is discretized into bins of width 3 mph over the range 0 to 120 mph, and launch angle is discretized into bins of width 3 degrees over the range -90 to 90 degrees. This results in a grid that balances resolution and sample size within each cell. For each cell, the expected outcome is defined as the average result across all batted balls with similar EV and LA, aggregated over multiple seasons. This empirical approach avoids reliance on parametric assumptions and reflects league-wide average behavior for comparable contact profiles.

This naturally leads to a residual-based representation of batted-ball outcomes. We therefore define the outcome residual at the level of individual batted balls. Let
$TB_i$ denote the total bases recorded for batted ball $i$, and let $(EV_i, LA_i)$
denote the corresponding exit velocity and launch angle. For each $(EV, LA)$ grid cell $g$, we define the expected total bases as the empirical mean
\[
\mu_g = \mathbb{E}(TB \mid EV, LA \in g),
\]
estimated by averaging $TB_i$ over all batted balls whose physical
characteristics fall within cell $g$. The total bases residual (TBR)  for batted ball $i$ is
then defined as
\begin{equation}
    R_i = TB_i - \mu_{g(i)},\label{eq:residual}
\end{equation}
where $g(i)$ denotes the grid cell containing $(EV_i, LA_i)$.

By construction, $R_i$ measures whether batted ball $i$ resulted in more or
fewer bases than would be expected given its contact quality. Positive values
of $R_i$ correspond to outcomes better than the league-wide average for
comparable contact profiles, while negative values indicate worse-than-average
outcomes. Conditional on exit velocity and launch angle, the expected outcome
$\mu_g$ does not depend on the identities of the batter or pitcher, as their
primary influence is expressed through contact quality. As a result, variation
in $R_i$ primarily reflects the combined effects of ballpark environment and
team-level defensive performance, with remaining variation attributable to
random factors not explicitly modeled. The defensive effects captured here are
restricted to batted-ball events and do not include aspects such as catcher
framing, blocking, or control of the running game.

\subsection{Simultaneous Estimation Model}\label{sec:simultaneous}

Our objective is to estimate ballpark effects and team-level defensive effects from the outcome residuals constructed in Section~\ref{sec:TBR}. We begin by describing the model
at the level of individual batted balls. Let $R_i$ denote the outcome residual for batted ball $i$, as defined in \eqref{eq:residual}. Let $p(i)\in\{1,\dots,30\}$ denote the ballpark in which batted ball
$i$ was fielded, and let $d(i)\in\{1,\dots,30\}$ denote the defensive team on the
field for that play. We assume the additive structure
\begin{equation}
R_i = \beta_0 + \beta^{\text{park}}_{p(i)} - \beta^{\text{def}}_{d(i)} + \varepsilon_i ,
\label{eq:ball_level_model}
\end{equation}
where $\beta^{\text{park}}_{p}$ captures the environmental effect of ballpark $p$
and $\beta^{\text{def}}_{d}$ captures the team-level defensive effect of defensive
team $d$. The defensive effect enters the model with a negative sign so that larger values of $\beta^{\mathrm{def}}_d$ correspond to stronger defense. Holding the ballpark effect fixed, a larger defensive effect lowers the expected TBR, indicating fewer total bases allowed than expected.

This additive specification follows from the decomposition of batted-ball
outcomes, where the residual $R_i$ captures the combined effects of ballpark
environment and team defense. Modeling these effects additively provides a
simple and interpretable framework that allows them to be estimated
simultaneously across all observations.
In the construction of the residual $R_i$ defined in Section~\ref{sec:TBR}, contact quality has been controlled for, so the model focuses on systematic variation attributable to ballpark and team defense.

All effects are modeled using categorical indicators. To ensure identifiability,
one ballpark effect and one defensive team effect are set to zero and treated as
reference categories. Throughout the analysis, we take the Atlanta Braves as the
reference defensive team and their home ballpark in each season as the reference ballpark.
Under this parameterization, the model includes an intercept, $29$ ballpark effects,
and $29$ defensive team effects, and estimated coefficients are interpreted relative
to the chosen reference categories.

For estimation, we aggregate observations with identical covariate values to reduce the
effective sample size. At the batted-ball level, each season contains on the order of
$10^5$ observations. Under model \eqref{eq:ball_level_model}, the covariates depend only
on the pair $(p(i), d(i))$, so many observations share the same row in the design matrix. Aggregating within each $(p,d)$ cell reduces the regression to at most $30\times30=900$ cell-level observations per season, and in practice substantially fewer because not every defensive
team appears in every ballpark within a season. This aggregation yields a much more
efficient estimation procedure.

Specifically, for each cell $(p,d)$, let $n_{pd}$ denote the number of batted balls
(defensive opportunities), and define the average TBR
{
\begin{equation}
y_{pd} = \frac{1}{n_{pd}} \sum_{i:\, p(i)=p,\, d(i)=d} R_i.
\label{eq:cell_mean}
\end{equation}
}

We then estimate
\begin{equation}
y_{pd} = \beta_0 + \beta^{\text{park}}_{p} - \beta^{\text{def}}_{d} + \varepsilon_{pd},
\label{eq:cell_level_model}
\end{equation}
using weighted least squares with weight $n_{pd}$.

This aggregation strategy does not alter the estimated coefficients. Because all observations within a given $(p,d)$ cell share the same row in the design matrix,
minimizing the batted-ball-level least squares objective is equivalent to minimizing
the weighted sum of squared deviations of the cell means (see, e.g., \citep[Sec.~9.2]{kmenta1986elements}).  Consequently, the weighted regression in \eqref{eq:cell_level_model} yields coefficient estimates that are numerically identical to ordinary least squares applied directly to the individual-level model in \eqref{eq:ball_level_model} \citep[Sec. 3.1.3]{angrist2009mostly}. 

\subsection{Presentation and Interpretation of Estimated Effects}

\subsubsection{Centering Effects at the League Average}

The model in Section~\ref{sec:simultaneous} is estimated under a reference-based parameterization,
in which one ballpark effect and one defensive team effect are set to zero for
identifiability. Let $\hat{\beta}^{\text{park}}_p$ and $\hat{\beta}^{\text{def}}_d$
denote the estimated ballpark and defensive effects obtained under this parameterization for a given season.

For interpretation, we re-center the estimated effects so that they are expressed relative to the league average. Specifically, let
\begin{equation}
\bar{\beta}^{\text{park}} = \frac{1}{30} \sum_{p=1}^{30} \hat{\beta}^{\text{park}}_p,
\qquad
\bar{\beta}^{\text{def}} = \frac{1}{30} \sum_{d=1}^{30} \hat{\beta}^{\text{def}}_d .
\label{eq:mean_effects}
\end{equation}
We define the centered effects as
\begin{equation}
\tilde{\beta}^{\text{park}}_p = \hat{\beta}^{\text{park}}_p - \bar{\beta}^{\text{park}},
\qquad
\tilde{\beta}^{\text{def}}_d = \hat{\beta}^{\text{def}}_d - \bar{\beta}^{\text{def}} .
\label{eq:centered_effects}
\end{equation}
To preserve the fitted values, the intercept is adjusted accordingly as
\begin{equation}
\tilde{\beta}_0 = \hat{\beta}_0 + \bar{\beta}^{\text{park}} - \bar{\beta}^{\text{def}} .
\label{eq:centered_intercept}
\end{equation}

This transformation represents a linear re-expression of the same fitted model. Under the centered parameterization, $\tilde{\beta}^{\text{park}}_p > 0$ indicates a ballpark effect above the league average for that season, while $\tilde{\beta}^{\text{park}}_p < 0$ indicates a below-average effect. We refer to $\tilde{\beta}^{\mathrm{park}}_p$ as the TBR-based ballpark effect and to $\tilde{\beta}^{\mathrm{def}}_d$ as Defensive Bases Saved (DBS). Both are expressed on the per-batted-ball TBR scale.

\subsubsection{A Standardized Index for Interpreting Effect Magnitudes}

After centering the estimated effects at the league average, the sign of an effect
indicates whether it is above or below average. However, the magnitude of the effect
is still difficult to interpret on its original scale. This issue also appears in
commonly used 100-based baseball statistics such as SLG+ or ERA+, which are defined
as ratios to the league average multiplied by 100. While such measures clearly show
whether performance is above or below average, they provide limited information
about how extreme a given value is within the overall distribution.

To provide a clearer sense of effect magnitude, we use a standardized scale based on
the z-score. For a given set of estimated effects $\hat{\beta}$ within a season, we define
\begin{equation}
z = \frac{\hat{\beta} - \bar{\beta}}{s_{\beta}},
\end{equation}
where $\bar{\beta}$ and $s_{\beta}$ denote the mean and standard deviation of the
estimated effects, respectively. A z-score of 1 indicates an effect one standard
deviation above the league average, while larger values indicate increasingly rare
outcomes within the distribution. Table~\ref{tab:ztable} summarizes the upper-tail
probabilities associated with common z-score values.

\begin{table}[h]
\small\sf\centering
\caption{Upper-tail probabilities for standard normal z-scores.\label{tab:ztable}}
\begin{adjustbox}{max width=\columnwidth}
\begin{tabular}{cccccc}
\toprule
z-score & 1 & 2 & 3 & 4 & 5 \\
\midrule
$\Pr(Z>z)$ & 15.87\% & 2.28\% & 0.135\% & 0.0032\% & 0.00003\% \\
\bottomrule
\end{tabular}
\end{adjustbox}
\end{table}

For consistency with existing baseball metrics, we map the z-score to a 100-based
index defined as
\begin{equation}\label{eq:index}
\text{Index} = 100 + 20 \times z.
\end{equation}
Under this scale, a value of 100 corresponds to the league average, 120 corresponds
to one standard deviation above average, and 140 corresponds to two standard deviations above average. This transformation preserves the relative ordering of the estimated effects while making their magnitude easier to interpret.
%For defensive effects, where lower values of the raw effect correspond to better
%performance, we reverse the direction of the scale and define
%\begin{equation}
%\text{Defense Index} = 100 - 20 \times z.
%\end{equation}
%Under this definition, higher index values consistently indicate better defensive
%performance. 
This standardized representation makes it easier to compare effect
magnitudes across teams, ballparks, and seasons.

We refer to the standardized ballpark index as the TBR-based Park Factor (TBR-PF$^{*}$) and to the standardized version of Defensive Bases Saved as DBS$^{*}$. The superscript asterisk denotes the z-score-based transformation to the 100-based scale in equation~\eqref{eq:index}, distinguishing these indices from conventional ratio-based ``plus'' metrics; it does not denote statistical significance. When existing metrics are transformed to the same scale for comparison, we denote them by MLB PF$^{*}$, Def$^{*}$, and OAA$^{*}$.

{
\subsubsection{Uncertainty Quantification}

The weighted least-squares fit provides an estimated covariance matrix for
the reference-based coefficients. Because the centered effects in
equation~\eqref{eq:centered_effects} are linear transformations of these
coefficients, their standard errors are computed as linear contrasts of the
coefficient covariance matrix. For a centered effect
$\tilde{\beta}$, the 95\% confidence interval on the TBR scale is
\[
\tilde{\beta}
\pm
1.96\,\mathrm{SE}(\tilde{\beta}).
\]
For TBR-PF$^{*}$ and DBS$^{*}$, the two endpoints are transformed to the
100-based scale as
\[
100+
20\,
\frac{
\tilde{\beta}
\pm
1.96\,\mathrm{SE}(\tilde{\beta})
}{
s_{\beta}
}.
\]
Here, the seasonal standard deviation $s_{\beta}$ is treated as a plug-in
scaling constant. These intervals quantify uncertainty in the
season-specific ballpark and defensive effects.
}

\section{Results and Discussion}
We present the results in three parts. We first examine the estimated ballpark effects, quantify their uncertainty, and compare them with observed home–away patterns. We then present team-level defensive effects, quantify their uncertainty, and compare them with existing metrics. Finally, we examine the estimated league-wide intercept over time.
\subsection{Estimated Ballpark Effects}\label{{sec:park_results}}

We estimate ballpark effects separately for each season using Statcast data
from 2015 to 2024. We first present the estimated ballpark effects obtained
from our model. Table~\ref{tab:park_beta} reports the centered estimates
$\tilde{\beta}^{\text{park}}_{p}$ as defined in equation~\eqref{eq:centered_effects}. Standard errors are computed from the weighted least-squares fit in each season. To facilitate
comparison, Table~\ref{tab:std_comparison_side_by_side} reports the corresponding standardized indices defined in equation~\eqref{eq:index}, which place MLB PF$^{*}$ and TBR-PF$^{*}$ on a comparable scale. Based on the ten-year arithmetic average, Coors Field (COL) is estimated to be the ballpark most conducive to total bases, whereas Busch Stadium (STL) is estimated to be the least conducive.

As a basic check, we examine whether the estimated effects align with
widely recognized hitter-friendly and pitcher-friendly ballparks. As
expected, hitter-friendly parks such as Coors Field (COL) and Great
American Ball Park (CIN) are estimated to have relatively large positive
effects. On the other hand, parks that are commonly regarded as
pitcher-friendly, such as Oracle Park (SF) and T-Mobile Park (SEA),
tend to have lower estimated values.

\begin{table*}[htbp]
\small\sf\centering
\caption{Estimated TBR-based ballpark effects ($\tilde{\beta}^{\text{park}}_p$), 2015--2024. Standard errors, shown in parentheses, are computed as linear contrasts of the weighted least-squares coefficient covariance matrix under the centered parameterization.\label{tab:park_beta}}
\begin{adjustbox}{max totalsize={\textwidth}{0.92\textheight}}
\begin{tabular}{lrrrrr}
\toprule
Team & 2015 & 2016 & 2017 & 2018 & 2019 \\
\midrule
ATL & -0.0450 (0.0149) & -0.0400 (0.0139) & -0.0026 (0.0125) & -0.0117 (0.0138) & -0.0099 (0.0143) \\
AZ & 0.0056 (0.0148) & 0.0285 (0.0140) & 0.0110 (0.0130) & 0.0017 (0.0140) & -0.0024 (0.0140) \\
BAL & 0.0254 (0.0149) & -0.0095 (0.0142) & -0.0120 (0.0126) & 0.0075 (0.0137) & 0.0194 (0.0139) \\
BOS & 0.0256 (0.0145) & 0.0536 (0.0138) & -0.0048 (0.0126) & -0.0200 (0.0137) & -0.0159 (0.0140) \\
CHC & -0.0004 (0.0155) & -0.0133 (0.0145) & 0.0062 (0.0130) & 0.0152 (0.0138) & -0.0024 (0.0143) \\
CIN & 0.0292 (0.0147) & 0.0400 (0.0142) & 0.0477 (0.0129) & 0.0652 (0.0137) & 0.0744 (0.0145) \\
CLE & 0.0084 (0.0150) & 0.0235 (0.0141) & 0.0015 (0.0129) & -0.0247 (0.0137) & 0.0281 (0.0144) \\
COL & 0.0655 (0.0145) & 0.0766 (0.0138) & 0.0926 (0.0127) & 0.0924 (0.0137) & 0.0853 (0.0137) \\
CWS & 0.0119 (0.0149) & 0.0229 (0.0142) & 0.0233 (0.0129) & 0.0072 (0.0140) & -0.0189 (0.0144) \\
DET & -0.0281 (0.0146) & -0.0387 (0.0142) & -0.0422 (0.0125) & -0.0447 (0.0137) & -0.0355 (0.0142) \\
HOU & 0.0716 (0.0154) & 0.0103 (0.0144) & 0.0230 (0.0127) & 0.0307 (0.0141) & 0.0667 (0.0141) \\
KC & -0.0038 (0.0145) & -0.0301 (0.0139) & -0.0251 (0.0126) & -0.0426 (0.0136) & -0.0344 (0.0140) \\
LAA & -0.0193 (0.0149) & -0.0259 (0.0139) & -0.0387 (0.0128) & -0.0264 (0.0139) & 0.0106 (0.0140) \\
LAD & -0.0086 (0.0152) & -0.0157 (0.0145) & -0.0087 (0.0132) & 0.0078 (0.0141) & 0.0297 (0.0143) \\
MIA & -0.0287 (0.0148) & -0.0272 (0.0142) & -0.0074 (0.0127) & -0.0419 (0.0136) & -0.0125 (0.0143) \\
MIL & 0.0157 (0.0150) & 0.0243 (0.0146) & 0.0330 (0.0131) & 0.0136 (0.0142) & 0.0055 (0.0145) \\
MIN & 0.0019 (0.0147) & 0.0041 (0.0139) & 0.0177 (0.0126) & -0.0076 (0.0138) & -0.0258 (0.0139) \\
NYM & -0.0157 (0.0150) & -0.0143 (0.0143) & -0.0121 (0.0128) & -0.0144 (0.0141) & 0.0058 (0.0142) \\
NYY & 0.0167 (0.0148) & 0.0180 (0.0142) & -0.0015 (0.0130) & 0.0117 (0.0140) & -0.0061 (0.0144) \\
OAK & -0.0070 (0.0145) & -0.0493 (0.0140) & -0.0065 (0.0129) & -0.0465 (0.0138) & -0.0200 (0.0140) \\
PHI & -0.0084 (0.0147) & -0.0013 (0.0144) & 0.0070 (0.0129) & 0.0326 (0.0142) & 0.0162 (0.0142) \\
PIT & -0.0392 (0.0148) & 0.0293 (0.0140) & -0.0024 (0.0126) & -0.0042 (0.0138) & -0.0004 (0.0138) \\
SD & 0.0138 (0.0152) & 0.0082 (0.0143) & 0.0044 (0.0132) & 0.0365 (0.0140) & -0.0320 (0.0146) \\
SEA & -0.0278 (0.0150) & -0.0129 (0.0142) & -0.0183 (0.0128) & -0.0159 (0.0139) & -0.0223 (0.0144) \\
SF & -0.0254 (0.0148) & -0.0023 (0.0139) & -0.0604 (0.0125) & 0.0029 (0.0138) & -0.0578 (0.0142) \\
STL & -0.0015 (0.0149) & -0.0387 (0.0141) & -0.0408 (0.0129) & -0.0479 (0.0138) & -0.0539 (0.0143) \\
TB & -0.0023 (0.0153) & -0.0101 (0.0146) & 0.0035 (0.0132) & -0.0064 (0.0140) & -0.0247 (0.0144) \\
TEX & 0.0025 (0.0146) & 0.0239 (0.0140) & 0.0092 (0.0128) & 0.0295 (0.0138) & 0.0106 (0.0141) \\
TOR & -0.0010 (0.0147) & 0.0006 (0.0142) & -0.0139 (0.0128) & -0.0134 (0.0138) & 0.0034 (0.0143) \\
WSH & -0.0316 (0.0151) & -0.0344 (0.0142) & 0.0173 (0.0128) & 0.0141 (0.0137) & 0.0191 (0.0141) \\
\midrule
Team & 2020 & 2021 & 2022 & 2023 & 2024 \\
\midrule
ATL & -0.0045 (0.0122) & -0.0032 (0.0139) & -0.0223 (0.0132) & 0.0015 (0.0127) & -0.0183 (0.0135) \\
AZ & -0.0013 (0.0208) & 0.0461 (0.0134) & 0.0008 (0.0128) & 0.0342 (0.0125) & 0.0060 (0.0128) \\
BAL & 0.0105 (0.0250) & 0.0139 (0.0135) & -0.0246 (0.0128) & -0.0313 (0.0128) & -0.0118 (0.0131) \\
BOS & 0.0869 (0.0245) & 0.0075 (0.0135) & 0.0257 (0.0127) & 0.0248 (0.0125) & 0.0226 (0.0133) \\
CHC & -0.0361 (0.0222) & 0.0244 (0.0141) & 0.0178 (0.0129) & 0.0187 (0.0127) & -0.0241 (0.0134) \\
CIN & 0.0517 (0.0227) & 0.0538 (0.0138) & 0.0636 (0.0130) & 0.0564 (0.0128) & 0.0595 (0.0134) \\
CLE & 0.0216 (0.0220) & 0.0188 (0.0139) & 0.0099 (0.0126) & -0.0153 (0.0125) & 0.0239 (0.0133) \\
COL & 0.0708 (0.0204) & 0.0603 (0.0133) & 0.0640 (0.0124) & 0.0446 (0.0125) & 0.0554 (0.0130) \\
CWS & 0.0238 (0.0217) & 0.0342 (0.0139) & -0.0060 (0.0126) & -0.0060 (0.0127) & -0.0143 (0.0132) \\
DET & -0.0224 (0.0223) & -0.0338 (0.0137) & -0.0269 (0.0129) & -0.0150 (0.0127) & -0.0053 (0.0133) \\
HOU & -0.0127 (0.0213) & 0.0103 (0.0134) & 0.0175 (0.0129) & 0.0003 (0.0126) & 0.0181 (0.0130) \\
KC & -0.0179 (0.0215) & -0.0182 (0.0134) & -0.0269 (0.0125) & -0.0312 (0.0124) & -0.0103 (0.0129) \\
LAA & -0.0079 (0.0208) & -0.0108 (0.0137) & 0.0063 (0.0131) & 0.0022 (0.0130) & 0.0143 (0.0133) \\
LAD & 0.0076 (0.0210) & 0.0142 (0.0140) & 0.0381 (0.0130) & -0.0008 (0.0127) & 0.0071 (0.0131) \\
MIA & -0.0116 (0.0245) & -0.0144 (0.0139) & -0.0215 (0.0130) & -0.0110 (0.0126) & -0.0045 (0.0129) \\
MIL & -0.0053 (0.0227) & 0.0065 (0.0141) & 0.0048 (0.0132) & 0.0094 (0.0130) & 0.0318 (0.0135) \\
MIN & -0.0099 (0.0221) & -0.0563 (0.0136) & -0.0331 (0.0128) & 0.0097 (0.0132) & 0.0195 (0.0130) \\
NYM & 0.0331 (0.0237) & -0.0180 (0.0142) & -0.0238 (0.0129) & -0.0177 (0.0129) & -0.0064 (0.0133) \\
NYY & 0.0184 (0.0253) & -0.0217 (0.0141) & -0.0063 (0.0131) & -0.0340 (0.0130) & -0.0215 (0.0132) \\
OAK & -0.0654 (0.0216) & -0.0400 (0.0137) & -0.0149 (0.0129) & -0.0141 (0.0129) & -0.0341 (0.0132) \\
PHI & -0.0164 (0.0236) & 0.0113 (0.0138) & 0.0002 (0.0129) & 0.0153 (0.0127) & 0.0277 (0.0131) \\
PIT & -0.0470 (0.0214) & -0.0012 (0.0136) & -0.0132 (0.0129) & -0.0328 (0.0128) & -0.0243 (0.0132) \\
SD & -0.0167 (0.0218) & -0.0147 (0.0138) & -0.0274 (0.0130) & -0.0091 (0.0128) & 0.0015 (0.0130) \\
SEA & -0.0222 (0.0219) & -0.0305 (0.0139) & 0.0074 (0.0131) & 0.0015 (0.0130) & -0.0445 (0.0139) \\
SF & 0.0114 (0.0209) & -0.0235 (0.0138) & -0.0158 (0.0130) & -0.0378 (0.0128) & -0.0256 (0.0132) \\
STL & -0.0363 (0.0226) & -0.0379 (0.0137) & -0.0196 (0.0126) & -0.0378 (0.0123) & -0.0195 (0.0130) \\
TB & 0.0406 (0.0260) & 0.0227 (0.0140) & 0.0190 (0.0130) & 0.0393 (0.0130) & -0.0035 (0.0134) \\
TEX & -0.0140 (0.0216) & -0.0105 (0.0136) & 0.0115 (0.0129) & 0.0385 (0.0126) & -0.0121 (0.0132) \\
TOR & -0.0009 (0.0254) & 0.0175 (0.0135) & 0.0161 (0.0127) & 0.0041 (0.0127) & 0.0068 (0.0131) \\
WSH & -0.0280 (0.0238) & -0.0067 (0.0136) & -0.0206 (0.0127) & -0.0069 (0.0122) & -0.0138 (0.0130) \\
\bottomrule
\end{tabular}
\end{adjustbox}
\end{table*}

\begin{table*}[htbp]
\small\sf\centering
\caption{Comparison of MLB PF$^{*}$ and TBR-PF$^{*}$ (2015--2024). Avg denotes the ten-year arithmetic average. Rows are ordered by the ten-year average of TBR-PF$^{*}$, from most to least conducive to total bases.\label{tab:std_comparison_side_by_side}}
\begin{adjustbox}{max width=\textwidth}
\begin{tabular}{lrrrrrrrrrrr|rrrrrrrrrrr}
\toprule
Team & \multicolumn{11}{c|}{MLB PF$^{*}$} & \multicolumn{11}{c}{TBR-PF$^{*}$} \\
 & 2015 & 2016 & 2017 & 2018 & 2019 & 2020 & 2021 & 2022 & 2023 & 2024 & Avg & 2015 & 2016 & 2017 & 2018 & 2019 & 2020 & 2021 & 2022 & 2023 & 2024 & Avg \\
\midrule
COL & 168 & 156 & 156 & 159 & 172 & 133 & 135 & 163 & 161 & 147 & 155.0 & 150 & 152 & 165 & 158 & 151 & 143 & 144 & 151 & 135 & 146 & 149.5 \\
CIN & 108 & 111 & 109 & 120 & 97 & 124 & 144 & 132 & 110 & 124 & 117.9 & 122 & 127 & 134 & 141 & 145 & 132 & 139 & 151 & 145 & 149 & 138.5 \\
HOU & 92 & 74 & 91 & 80 & 109 & 61 & 101 & 95 & 100 & 114 & 91.7 & 154 & 107 & 116 & 119 & 140 & 92 & 107 & 114 & 100 & 115 & 116.4 \\
BOS & 128 & 134 & 100 & 112 & 113 & 138 & 131 & 136 & 137 & 110 & 123.9 & 119 & 136 & 97 & 87 & 90 & 153 & 105 & 121 & 120 & 119 & 114.7 \\
MIL & 112 & 100 & 105 & 96 & 97 & 109 & 85 & 91 & 91 & 82 & 96.8 & 112 & 116 & 123 & 109 & 103 & 97 & 105 & 104 & 107 & 126 & 110.2 \\
AZ & 120 & 137 & 123 & 100 & 97 & 109 & 118 & 91 & 96 & 128 & 111.9 & 104 & 119 & 108 & 101 & 99 & 99 & 133 & 101 & 127 & 105 & 109.6 \\
CLE & 116 & 119 & 81 & 112 & 89 & 100 & 101 & 82 & 73 & 110 & 98.3 & 106 & 116 & 101 & 85 & 117 & 113 & 114 & 108 & 88 & 120 & 106.8 \\
PHI & 112 & 89 & 109 & 92 & 113 & 114 & 106 & 113 & 96 & 100 & 104.4 & 94 & 99 & 105 & 120 & 110 & 90 & 108 & 100 & 112 & 123 & 106.1 \\
TEX & 128 & 122 & 132 & 143 & 128 & 100 & 89 & 109 & 128 & 77 & 115.6 & 102 & 116 & 106 & 118 & 106 & 91 & 92 & 109 & 131 & 90 & 106.1 \\
TB & 80 & 81 & 77 & 80 & 73 & 76 & 76 & 82 & 91 & 82 & 79.8 & 98 & 93 & 102 & 96 & 85 & 125 & 116 & 115 & 131 & 97 & 105.8 \\
CWS & 88 & 96 & 105 & 92 & 97 & 109 & 106 & 104 & 96 & 96 & 98.9 & 109 & 115 & 116 & 104 & 89 & 115 & 125 & 95 & 95 & 88 & 105.1 \\
LAD & 80 & 78 & 77 & 88 & 89 & 109 & 97 & 104 & 96 & 100 & 91.8 & 93 & 89 & 94 & 105 & 118 & 105 & 110 & 130 & 99 & 106 & 104.9 \\
TOR & 96 & 104 & 95 & 108 & 101 & 104 & 80 & 113 & 87 & 100 & 98.8 & 99 & 100 & 90 & 92 & 102 & 99 & 113 & 113 & 103 & 106 & 101.7 \\
CHC & 100 & 78 & 114 & 108 & 97 & 80 & 114 & 95 & 105 & 59 & 95.0 & 100 & 91 & 104 & 110 & 99 & 78 & 118 & 114 & 115 & 80 & 100.9 \\
BAL & 116 & 104 & 114 & 108 & 121 & 71 & 135 & 95 & 82 & 110 & 105.6 & 119 & 94 & 92 & 105 & 112 & 106 & 110 & 80 & 75 & 90 & 98.3 \\
SD & 96 & 93 & 72 & 100 & 81 & 95 & 89 & 63 & 82 & 96 & 86.7 & 110 & 106 & 103 & 123 & 81 & 90 & 89 & 78 & 93 & 101 & 97.4 \\
NYY & 104 & 96 & 95 & 116 & 89 & 104 & 89 & 91 & 96 & 114 & 99.4 & 113 & 112 & 99 & 107 & 96 & 111 & 84 & 95 & 73 & 82 & 97.2 \\
MIN & 108 & 115 & 119 & 108 & 101 & 71 & 106 & 100 & 100 & 124 & 105.2 & 101 & 103 & 112 & 95 & 84 & 94 & 59 & 74 & 108 & 116 & 94.6 \\
LAA & 76 & 89 & 77 & 88 & 101 & 138 & 106 & 100 & 100 & 105 & 98.0 & 85 & 82 & 73 & 83 & 106 & 95 & 92 & 105 & 102 & 112 & 93.5 \\
NYM & 72 & 85 & 81 & 65 & 89 & 104 & 85 & 82 & 87 & 96 & 84.6 & 88 & 90 & 91 & 91 & 103 & 120 & 87 & 81 & 86 & 95 & 93.2 \\
WSH & 100 & 85 & 109 & 124 & 121 & 85 & 118 & 109 & 114 & 96 & 106.1 & 76 & 77 & 112 & 109 & 111 & 83 & 95 & 84 & 95 & 89 & 93.1 \\
PIT & 88 & 115 & 91 & 100 & 109 & 90 & 110 & 104 & 91 & 110 & 100.8 & 70 & 120 & 98 & 97 & 100 & 71 & 99 & 89 & 74 & 80 & 89.8 \\
ATL & 88 & 96 & 105 & 92 & 117 & 104 & 101 & 95 & 114 & 96 & 100.8 & 66 & 73 & 98 & 93 & 94 & 97 & 98 & 82 & 101 & 85 & 88.7 \\
MIA & 84 & 78 & 91 & 69 & 85 & 90 & 76 & 100 & 105 & 119 & 89.7 & 78 & 82 & 95 & 74 & 92 & 93 & 90 & 83 & 91 & 96 & 87.4 \\
SEA & 88 & 96 & 86 & 84 & 85 & 76 & 68 & 68 & 68 & 49 & 76.8 & 79 & 91 & 87 & 90 & 87 & 86 & 78 & 106 & 101 & 63 & 86.8 \\
KC & 96 & 96 & 86 & 104 & 101 & 109 & 110 & 122 & 128 & 110 & 106.2 & 97 & 80 & 82 & 73 & 79 & 89 & 87 & 79 & 75 & 91 & 83.2 \\
SF & 76 & 100 & 68 & 92 & 65 & 114 & 93 & 100 & 73 & 82 & 86.3 & 81 & 98 & 57 & 102 & 65 & 107 & 83 & 87 & 70 & 79 & 82.9 \\
DET & 100 & 111 & 128 & 88 & 105 & 109 & 85 & 86 & 96 & 86 & 99.4 & 79 & 74 & 70 & 72 & 79 & 86 & 76 & 79 & 88 & 96 & 79.9 \\
OAK & 84 & 70 & 114 & 80 & 85 & 71 & 76 & 86 & 82 & 91 & 83.9 & 95 & 67 & 95 & 71 & 88 & 60 & 71 & 88 & 89 & 72 & 79.6 \\
STL & 92 & 93 & 91 & 88 & 81 & 100 & 72 & 91 & 114 & 91 & 91.3 & 99 & 74 & 71 & 70 & 68 & 78 & 73 & 84 & 70 & 84 & 77.1 \\
\bottomrule
\end{tabular}
\end{adjustbox}
\end{table*}

Because ballpark effects are primarily determined by physical
characteristics of the stadium, they are expected to remain relatively
stable over time. We therefore examine the temporal variability of the
estimated effects. Table~\ref{tab:team_comparison} reports the standard deviation of the estimated
effects for each team over the ten seasons. On average, our estimates
exhibit slightly lower variability than the corresponding MLB official
park factors (12.51~vs~13.08).

The uncertainty estimates provide additional context for interpreting this
year-to-year variation. Across the 300 park-season estimates, the average
half-width of the 95\% confidence intervals on the standardized scale is
19.97 index points, with a standard deviation of 2.94. Thus, even with several
thousand batted-ball observations in a season, the precision attainable for a
single-season park effect remains limited. This helps explain why estimated
park factors may fluctuate across seasons even when the physical
characteristics of the ballpark remain largely unchanged: part of the observed
variation may reflect sampling variability rather than a genuine change in the
underlying park environment. As shown in
Appendix~\ref{app:park_ci}, this level of uncertainty is consistent with the
observed variability of individual TBR values and the number of batted balls
available in a single season. The appendix also reports the full confidence
intervals and the distribution of their half-widths.

This uncertainty does not diminish the usefulness of the single-season
estimates. The point estimate provides our best estimate of the ballpark
effect for that season, while the confidence interval describes the precision
of the estimate. A season-specific park effect may differ from the long-run
average because the effect of the ballpark can depend on the distribution and
direction of batted balls and on other season-specific conditions. The case
studies below show that our estimates are often more consistent with the
observed home--away patterns for the corresponding season than the official
MLB park factors.

We also observe that the largest discrepancies tend to occur in ballparks
such as Fenway Park (BOS), Yankee Stadium (NYY), and PNC Park (PIT), which are known to exhibit strong
asymmetries affecting left- and right-handed batters. Because our model
does not explicitly incorporate the direction of batted balls, it may be
less able to capture such asymmetric effects. Similar patterns are observed
for several other teams with relatively large differences.

\begin{table}[htbp]
\small\sf\centering
\caption{Ten-Year Standard Deviation of MLB PF$^{*}$ and TBR-PF$^{*}$, 2015--2024.\label{tab:team_comparison}}
\begin{adjustbox}{max width=\columnwidth}
\begin{tabular}{lcc@{\hspace{3ex}}lcc}
\toprule
\textbf{Team} & \textbf{MLB PF$^{*}$} & \textbf{TBR-PF$^{*}$} & \textbf{Team} & \textbf{MLB PF$^{*}$} & \textbf{TBR-PF$^{*}$} \\ 
\midrule
ATL & 9.27 & 11.85 & MIL & 10.06 & 9.30 \\ 
AZ & 15.78 & 12.48 & MIN & 14.44 & 17.82 \\ 
BAL & 18.78 & 14.35 & NYM & 11.12 & 11.15 \\ 
BOS & 13.91 & 20.49 & NYY & 9.91 & 13.92 \\ 
CHC & 17.85 & 14.17 & OAK & 12.44 & 12.76 \\ 
CIN & 13.51 & 9.56 & PHI & 9.47 & 10.76 \\ 
CLE & 16.24 & 12.15 & PIT & 10.01 & 15.93 \\ 
COL & 13.03 & 8.28 & SD & 11.87 & 13.85 \\ 
CWS & 6.72 & 12.81 & SEA & 13.80 & 12.02 \\ 
DET & 14.05 & 7.97 & SF & 16.12 & 16.15 \\ 
HOU & 16.16 & 18.32 & STL & 11.05 & 9.61 \\ 
KC & 12.57 & 7.61 & TB & 5.03 & 15.03 \\ 
LAA & 17.75 & 12.57 & TEX & 21.33 & 13.08 \\ 
LAD & 11.37 & 12.37 & TOR & 9.82 & 7.52 \\ 
MIA & 14.99 & 7.69 & WSH & 14.05 & 13.85 \\ 
\midrule
\textbf{Average} & \textbf{13.08} & \textbf{12.51} & & & \\ 
\bottomrule
\end{tabular}
\end{adjustbox}
\end{table}

To further examine the differences between our estimates and the official MLB park factors, we focus on several cases with the largest discrepancies, including Comerica Park (DET, 2017), Tropicana Field (TB, 2020--2023), Minute Maid Park (HOU, 2015), and Target Field (MIN, 2021). For each case, we compare offensive performance at home and away using the Total Bases Residual (TBR) framework. Table~\ref{tab:park_effect_case_study} reports the average values of Team TBR and Opponent TBR in home and away games; the corresponding values for all ballparks are provided in Appendix~\ref{app:home_away_residuals}.

\begin{table}[htbp]
\small\sf\centering
\caption{Home and away TBR for selected ballpark case studies, with corresponding MLB PF$^{*}$ and TBR-PF$^{*}$ values.\label{tab:park_effect_case_study}}
\begin{adjustbox}{max width=\columnwidth}
\begin{tabular}{lcccccc}
\toprule
\multirow{2}{*}{Team (Season)} & \multicolumn{2}{c}{Opponent TBR} & \multicolumn{2}{c}{Team TBR} & \multicolumn{2}{c}{Park Effect} \\
\cmidrule(lr){2-3} \cmidrule(lr){4-5} \cmidrule(lr){6-7}
 & Home & Away & Home & Away & MLB PF$^{*}$ & TBR-PF$^{*}$  \\
\midrule
DET (2017) &  0.003 &  0.064 &  0.002 & -0.004 & 128 & 70 \\
HOU (2015) &  0.035 &  0.016 &  0.116 &  0.002 & 92  & 154 \\
MIN (2021) & -0.056 &  0.013 & -0.072 & -0.051 & 106 & 59 \\
TB (2020)  &  0.028 &  0.006 &  0.058 &  0.025 & 76  & 125 \\
TB (2021)  & -0.048 & -0.044 &  0.026 &  0.009 & 76  & 116 \\
TB (2022)  & -0.042 & -0.040 & -0.011 & -0.024 & 82  & 115 \\
TB (2023)  & -0.021 & -0.045 &  0.026 &  0.009 & 91  & 131 \\
\bottomrule
\end{tabular}
\end{adjustbox}
\end{table}

We begin with Comerica Park (DET, 2017), which exhibits the largest discrepancy between the two measures. For visiting teams, the average TBR is 0.003 at Comerica Park and 0.064 in away games, indicating lower offensive production at this venue. Detroit, by contrast, records average TBR values of 0.002 at home and -0.004 on the road, so the home–away difference for the team is small and goes in the opposite direction. Because the opponent pattern is much stronger, the overall evidence is still more consistent with a pitcher-friendly environment, supporting our estimate below 100 (70) rather than the higher value implied by the official MLB metric (128).

A contrasting example is Minute Maid Park (HOU, 2015). In this case, Houston records an average TBR of 0.116 at home compared to 0.002 on the road, indicating substantially higher offensive production at its home ballpark. Visiting teams show a similar pattern, with average TBR values of 0.035 at Minute Maid Park and 0.016 in away games. This consistent increase in offensive output for both Houston and its opponents suggests a hitter-friendly environment, supporting our estimate above 100 (154) rather than the lower value implied by the official MLB metric (92). Similar patterns are observed in the remaining cases, where the direction of the home–away differences for both the focal team and its opponents is generally consistent with whether the estimated park effect is above or below the league average, providing additional support for the proposed method.

Taken together, these examples illustrate a systematic difference between the two approaches. The MLB park factors are based directly on observed offensive outcomes, which may reflect not only the ballpark environment but also differences in team defense and contact quality. In contrast, our model-based estimates explicitly control for these factors, allowing the environmental component to be more clearly isolated. This distinction provides a possible explanation for why our estimates appear more consistent with the observed home–away patterns in these cases.

\subsection{Estimation of Team Defensive Effects}

In the previous subsection, we presented the estimated ballpark effects obtained from the model in \eqref{eq:ball_level_model}. As part of the same estimation, the model also yields team-level defensive effects, which we now present. Table~\ref{tab:defense_beta} reports Defensive Bases Saved (DBS), the centered estimates $\tilde{\beta}^{\text{def}}_{d}$ as defined in equation~\eqref{eq:centered_effects}. Unlike ballpark effects, team defensive effects are not expected to exhibit strong year-to-year stability due to roster changes. Therefore, we do not examine stability and proceed directly to compare our estimates with existing defensive metrics. Specifically, we compare our estimates with Defensive Runs Above Average (Def) and Outs Above Average (OAA), which represent two widely used approaches to quantifying defensive performance.

\begin{table*}[htbp]
\small\sf\centering
\caption{Defensive Bases Saved (DBS) ($\tilde{\beta}^{\text{def}}_d$), 2015--2024. Standard errors, shown in parentheses, are computed as linear contrasts of the weighted least-squares coefficient covariance matrix under the centered parameterization.\label{tab:defense_beta}}
\begin{adjustbox}{max totalsize={\textwidth}{0.92\textheight}}
\begin{tabular}{lrrrrr}
\toprule
Team & 2015 & 2016 & 2017 & 2018 & 2019 \\
\midrule
ATL & -0.0272 (0.0148) & -0.0175 (0.0140) & -0.0122 (0.0126) & 0.0199 (0.0141) & 0.0156 (0.0142) \\
AZ & -0.0008 (0.0147) & 0.0066 (0.0141) & 0.0174 (0.0131) & 0.0417 (0.0140) & 0.0343 (0.0141) \\
BAL & -0.0098 (0.0150) & -0.0212 (0.0141) & -0.0088 (0.0127) & -0.0295 (0.0135) & -0.0051 (0.0138) \\
BOS & 0.0009 (0.0146) & 0.0457 (0.0143) & 0.0136 (0.0130) & -0.0113 (0.0140) & -0.0176 (0.0142) \\
CHC & -0.0015 (0.0153) & 0.0329 (0.0146) & 0.0216 (0.0131) & 0.0377 (0.0138) & 0.0207 (0.0142) \\
CIN & -0.0055 (0.0147) & -0.0039 (0.0141) & -0.0311 (0.0127) & 0.0145 (0.0135) & 0.0050 (0.0146) \\
CLE & 0.0206 (0.0153) & 0.0095 (0.0143) & 0.0166 (0.0133) & -0.0424 (0.0140) & 0.0133 (0.0144) \\
COL & -0.0206 (0.0146) & 0.0134 (0.0140) & 0.0093 (0.0128) & 0.0140 (0.0139) & 0.0230 (0.0138) \\
CWS & -0.0263 (0.0150) & 0.0037 (0.0141) & 0.0096 (0.0128) & 0.0117 (0.0137) & -0.0123 (0.0142) \\
DET & -0.0366 (0.0146) & -0.0242 (0.0141) & -0.0156 (0.0126) & -0.0302 (0.0137) & -0.0292 (0.0142) \\
HOU & 0.0314 (0.0151) & -0.0254 (0.0142) & -0.0178 (0.0131) & 0.0031 (0.0145) & 0.0113 (0.0146) \\
KC & 0.0211 (0.0147) & 0.0114 (0.0142) & -0.0099 (0.0126) & -0.0190 (0.0135) & 0.0032 (0.0140) \\
LAA & 0.0065 (0.0148) & -0.0012 (0.0139) & 0.0108 (0.0129) & -0.0061 (0.0140) & -0.0128 (0.0141) \\
LAD & -0.0125 (0.0153) & 0.0010 (0.0145) & -0.0076 (0.0133) & -0.0140 (0.0141) & 0.0136 (0.0145) \\
MIA & 0.0140 (0.0149) & -0.0016 (0.0143) & 0.0107 (0.0126) & -0.0131 (0.0137) & 0.0003 (0.0143) \\
MIL & -0.0130 (0.0150) & -0.0041 (0.0141) & 0.0090 (0.0130) & 0.0213 (0.0140) & -0.0067 (0.0141) \\
MIN & 0.0321 (0.0145) & -0.0006 (0.0139) & 0.0206 (0.0126) & -0.0143 (0.0138) & 0.0010 (0.0141) \\
NYM & -0.0190 (0.0150) & 0.0011 (0.0142) & -0.0263 (0.0126) & -0.0222 (0.0138) & -0.0199 (0.0141) \\
NYY & -0.0074 (0.0149) & -0.0129 (0.0143) & 0.0204 (0.0131) & -0.0049 (0.0142) & -0.0293 (0.0143) \\
OAK & 0.0237 (0.0147) & -0.0064 (0.0140) & 0.0135 (0.0127) & 0.0187 (0.0137) & 0.0230 (0.0140) \\
PHI & -0.0189 (0.0145) & -0.0211 (0.0141) & -0.0070 (0.0127) & -0.0154 (0.0139) & 0.0075 (0.0141) \\
PIT & 0.0096 (0.0148) & -0.0248 (0.0139) & -0.0051 (0.0127) & 0.0094 (0.0137) & -0.0396 (0.0140) \\
SD & -0.0253 (0.0151) & 0.0131 (0.0142) & -0.0208 (0.0129) & 0.0115 (0.0139) & -0.0146 (0.0144) \\
SEA & -0.0135 (0.0148) & -0.0054 (0.0141) & 0.0080 (0.0126) & 0.0085 (0.0137) & -0.0163 (0.0139) \\
SF & 0.0025 (0.0148) & 0.0190 (0.0141) & -0.0098 (0.0125) & 0.0273 (0.0137) & 0.0104 (0.0141) \\
STL & 0.0104 (0.0151) & -0.0197 (0.0141) & -0.0198 (0.0129) & 0.0044 (0.0138) & 0.0016 (0.0144) \\
TB & 0.0415 (0.0151) & 0.0081 (0.0143) & 0.0111 (0.0129) & 0.0159 (0.0140) & -0.0135 (0.0145) \\
TEX & 0.0124 (0.0146) & 0.0239 (0.0140) & 0.0164 (0.0127) & 0.0013 (0.0136) & -0.0009 (0.0142) \\
TOR & 0.0161 (0.0147) & 0.0266 (0.0142) & -0.0287 (0.0127) & -0.0441 (0.0137) & 0.0022 (0.0141) \\
WSH & -0.0050 (0.0151) & -0.0261 (0.0145) & 0.0119 (0.0130) & 0.0056 (0.0139) & 0.0317 (0.0144) \\
\midrule
Team & 2020 & 2021 & 2022 & 2023 & 2024 \\
\midrule
ATL & 0.0013 (0.0122) & 0.0225 (0.0139) & -0.0055 (0.0132) & 0.0078 (0.0128) & -0.0194 (0.0136) \\
AZ & -0.0082 (0.0211) & -0.0042 (0.0134) & 0.0166 (0.0127) & 0.0205 (0.0126) & -0.0137 (0.0129) \\
BAL & 0.0196 (0.0252) & -0.0002 (0.0134) & 0.0003 (0.0127) & 0.0156 (0.0128) & 0.0063 (0.0132) \\
BOS & 0.0262 (0.0248) & -0.0244 (0.0138) & 0.0139 (0.0128) & -0.0171 (0.0127) & 0.0141 (0.0131) \\
CHC & -0.0078 (0.0218) & -0.0084 (0.0138) & -0.0012 (0.0128) & 0.0232 (0.0127) & 0.0051 (0.0131) \\
CIN & -0.0144 (0.0229) & -0.0212 (0.0138) & -0.0181 (0.0130) & -0.0084 (0.0126) & 0.0070 (0.0133) \\
CLE & 0.0187 (0.0222) & 0.0216 (0.0139) & 0.0413 (0.0128) & 0.0228 (0.0125) & 0.0117 (0.0134) \\
COL & 0.0256 (0.0204) & 0.0021 (0.0136) & 0.0086 (0.0126) & -0.0077 (0.0124) & 0.0007 (0.0127) \\
CWS & 0.0445 (0.0216) & 0.0047 (0.0141) & -0.0172 (0.0129) & -0.0174 (0.0128) & -0.0242 (0.0132) \\
DET & -0.0056 (0.0215) & 0.0060 (0.0135) & 0.0219 (0.0128) & 0.0159 (0.0127) & -0.0021 (0.0132) \\
HOU & -0.0138 (0.0211) & 0.0150 (0.0137) & 0.0309 (0.0131) & 0.0162 (0.0128) & -0.0030 (0.0134) \\
KC & 0.0149 (0.0218) & 0.0311 (0.0137) & -0.0097 (0.0126) & -0.0255 (0.0126) & 0.0171 (0.0132) \\
LAA & -0.0133 (0.0212) & -0.0290 (0.0137) & 0.0132 (0.0130) & -0.0115 (0.0128) & -0.0011 (0.0131) \\
LAD & 0.0337 (0.0211) & 0.0184 (0.0141) & 0.0216 (0.0131) & 0.0079 (0.0127) & 0.0088 (0.0132) \\
MIA & -0.0151 (0.0235) & -0.0051 (0.0138) & -0.0398 (0.0131) & -0.0239 (0.0128) & -0.0088 (0.0130) \\
MIL & -0.0375 (0.0225) & 0.0109 (0.0143) & -0.0131 (0.0131) & 0.0171 (0.0129) & 0.0386 (0.0132) \\
MIN & 0.0249 (0.0220) & -0.0247 (0.0136) & -0.0142 (0.0128) & -0.0093 (0.0130) & -0.0017 (0.0132) \\
NYM & -0.0100 (0.0241) & -0.0001 (0.0140) & -0.0153 (0.0131) & -0.0033 (0.0128) & 0.0095 (0.0134) \\
NYY & -0.0000 (0.0259) & -0.0202 (0.0139) & -0.0014 (0.0131) & -0.0102 (0.0129) & -0.0118 (0.0133) \\
OAK & -0.0200 (0.0216) & 0.0010 (0.0136) & -0.0072 (0.0127) & -0.0276 (0.0126) & -0.0253 (0.0130) \\
PHI & -0.0672 (0.0242) & -0.0376 (0.0138) & -0.0207 (0.0130) & -0.0026 (0.0127) & -0.0003 (0.0132) \\
PIT & -0.0219 (0.0218) & -0.0168 (0.0137) & -0.0002 (0.0128) & -0.0053 (0.0128) & -0.0095 (0.0132) \\
SD & 0.0371 (0.0217) & 0.0058 (0.0139) & -0.0049 (0.0129) & -0.0048 (0.0128) & 0.0030 (0.0132) \\
SEA & -0.0246 (0.0214) & -0.0079 (0.0136) & 0.0048 (0.0129) & 0.0060 (0.0128) & -0.0104 (0.0134) \\
SF & 0.0201 (0.0212) & 0.0111 (0.0137) & -0.0302 (0.0128) & -0.0251 (0.0126) & -0.0149 (0.0133) \\
STL & 0.0302 (0.0232) & 0.0311 (0.0136) & 0.0105 (0.0126) & -0.0237 (0.0123) & -0.0077 (0.0130) \\
TB & 0.0147 (0.0255) & 0.0354 (0.0136) & 0.0170 (0.0129) & 0.0266 (0.0129) & 0.0140 (0.0132) \\
TEX & -0.0180 (0.0211) & -0.0097 (0.0135) & 0.0025 (0.0128) & 0.0186 (0.0127) & 0.0073 (0.0131) \\
TOR & -0.0102 (0.0251) & 0.0017 (0.0139) & 0.0227 (0.0128) & 0.0102 (0.0128) & 0.0154 (0.0132) \\
WSH & -0.0239 (0.0234) & -0.0089 (0.0136) & -0.0273 (0.0126) & 0.0152 (0.0124) & -0.0045 (0.0130) \\
\bottomrule
\end{tabular}
\end{adjustbox}
\end{table*}

\begin{table*}[htbp]
\small\sf\centering
\caption{Comparative View of Def$^{*}$ and DBS$^{*}$ (2015--2024). Avg denotes the ten-year arithmetic average. Rows are ordered by the ten-year average of DBS$^{*}$, from strongest to weakest team defense.\label{tab:std_defense_comparison}}
\begin{adjustbox}{max width=\textwidth}
\begin{tabular}{lrrrrrrrrrrr|rrrrrrrrrrr}
\toprule
Team & \multicolumn{11}{c|}{Def$^{*}$} & \multicolumn{11}{c}{DBS$^{*}$} \\
 & 2015 & 2016 & 2017 & 2018 & 2019 & 2020 & 2021 & 2022 & 2023 & 2024 & Avg & 2015 & 2016 & 2017 & 2018 & 2019 & 2020 & 2021 & 2022 & 2023 & 2024 & Avg \\
\midrule
TB & 120 & 96 & 108 & 98 & 112 & 108 & 123 & 102 & 93 & 103 & 106.3 & 143 & 109 & 114 & 115 & 85 & 112 & 139 & 118 & 132 & 121 & 118.8 \\
CLE & 112 & 121 & 137 & 102 & 141 & 114 & 115 & 119 & 106 & 127 & 119.4 & 121 & 111 & 121 & 60 & 115 & 115 & 124 & 144 & 127 & 117 & 115.5 \\
CHC & 125 & 148 & 105 & 113 & 113 & 127 & 104 & 84 & 109 & 111 & 113.9 & 98 & 136 & 127 & 135 & 123 & 94 & 91 & 99 & 128 & 108 & 113.9 \\
AZ & 99 & 80 & 97 & 139 & 130 & 120 & 86 & 127 & 115 & 117 & 111.0 & 99 & 107 & 122 & 139 & 138 & 93 & 95 & 118 & 125 & 80 & 111.6 \\
LAD & 122 & 116 & 112 & 111 & 117 & 109 & 98 & 110 & 101 & 91 & 108.7 & 87 & 101 & 91 & 87 & 115 & 127 & 120 & 123 & 109 & 113 & 107.3 \\
COL & 63 & 120 & 103 & 102 & 100 & 96 & 108 & 89 & 100 & 97 & 97.8 & 79 & 115 & 112 & 113 & 126 & 121 & 102 & 109 & 91 & 101 & 106.9 \\
TEX & 113 & 110 & 93 & 85 & 80 & 71 & 138 & 102 & 126 & 116 & 103.4 & 113 & 126 & 120 & 101 & 99 & 85 & 89 & 103 & 122 & 111 & 106.9 \\
MIL & 89 & 85 & 119 & 135 & 115 & 124 & 97 & 110 & 142 & 124 & 114.0 & 87 & 96 & 111 & 120 & 93 & 70 & 112 & 86 & 120 & 157 & 105.2 \\
HOU & 95 & 134 & 110 & 139 & 125 & 103 & 127 & 130 & 98 & 102 & 116.3 & 132 & 72 & 78 & 103 & 113 & 89 & 116 & 133 & 119 & 96 & 105.1 \\
BOS & 87 & 104 & 126 & 130 & 121 & 100 & 90 & 93 & 58 & 83 & 99.2 & 101 & 150 & 117 & 89 & 80 & 121 & 73 & 115 & 79 & 121 & 104.6 \\
KC & 128 & 112 & 100 & 84 & 92 & 117 & 109 & 79 & 120 & 129 & 107.0 & 122 & 113 & 88 & 82 & 104 & 112 & 134 & 90 & 69 & 125 & 103.9 \\
TOR & 121 & 110 & 73 & 84 & 98 & 79 & 107 & 119 & 119 & 130 & 104.0 & 117 & 129 & 64 & 59 & 102 & 92 & 102 & 124 & 112 & 123 & 102.4 \\
MIN & 113 & 79 & 146 & 99 & 92 & 112 & 113 & 97 & 96 & 105 & 105.2 & 133 & 99 & 126 & 87 & 101 & 120 & 73 & 85 & 89 & 97 & 101.0 \\
STL & 118 & 93 & 93 & 72 & 109 & 120 & 134 & 121 & 84 & 98 & 104.2 & 111 & 78 & 75 & 104 & 102 & 124 & 134 & 111 & 72 & 89 & 100.0 \\
SF & 126 & 123 & 82 & 86 & 112 & 98 & 124 & 63 & 120 & 107 & 104.1 & 103 & 121 & 88 & 126 & 112 & 116 & 112 & 68 & 70 & 78 & 99.4 \\
SD & 99 & 68 & 89 & 105 & 95 & 126 & 106 & 102 & 114 & 93 & 99.7 & 74 & 114 & 74 & 111 & 84 & 130 & 106 & 95 & 94 & 104 & 98.6 \\
OAK & 55 & 57 & 74 & 105 & 95 & 93 & 120 & 95 & 71 & 60 & 82.5 & 124 & 93 & 117 & 118 & 126 & 84 & 101 & 92 & 67 & 62 & 98.4 \\
ATL & 89 & 95 & 84 & 123 & 107 & 122 & 102 & 103 & 110 & 101 & 103.6 & 72 & 81 & 85 & 119 & 117 & 101 & 125 & 94 & 109 & 71 & 97.4 \\
WSH & 76 & 103 & 81 & 93 & 107 & 71 & 82 & 52 & 82 & 76 & 82.3 & 95 & 71 & 115 & 105 & 135 & 81 & 90 & 71 & 118 & 93 & 97.4 \\
BAL & 108 & 104 & 83 & 66 & 53 & 67 & 56 & 112 & 100 & 89 & 83.8 & 90 & 77 & 89 & 72 & 94 & 116 & 100 & 100 & 119 & 109 & 96.6 \\
LAA & 117 & 100 & 134 & 112 & 96 & 63 & 82 & 86 & 75 & 71 & 93.6 & 107 & 99 & 113 & 94 & 86 & 89 & 68 & 114 & 86 & 98 & 95.4 \\
CWS & 94 & 88 & 107 & 79 & 74 & 129 & 72 & 91 & 80 & 55 & 86.9 & 73 & 104 & 112 & 111 & 86 & 136 & 105 & 81 & 79 & 64 & 95.1 \\
SEA & 100 & 86 & 117 & 95 & 73 & 91 & 79 & 100 & 117 & 98 & 95.6 & 86 & 94 & 110 & 108 & 82 & 80 & 91 & 105 & 107 & 85 & 94.8 \\
CIN & 82 & 74 & 106 & 91 & 121 & 91 & 73 & 80 & 64 & 84 & 86.6 & 94 & 96 & 61 & 114 & 106 & 88 & 77 & 81 & 90 & 110 & 91.7 \\
NYY & 80 & 114 & 115 & 92 & 80 & 89 & 78 & 150 & 115 & 128 & 104.1 & 92 & 86 & 125 & 95 & 67 & 100 & 78 & 98 & 88 & 82 & 91.1 \\
MIA & 95 & 112 & 96 & 122 & 87 & 74 & 99 & 101 & 77 & 86 & 94.9 & 114 & 98 & 113 & 88 & 100 & 88 & 94 & 57 & 71 & 87 & 91.0 \\
DET & 83 & 77 & 73 & 79 & 78 & 99 & 77 & 100 & 102 & 126 & 89.4 & 62 & 73 & 81 & 72 & 67 & 96 & 107 & 124 & 119 & 97 & 89.8 \\
NYM & 113 & 98 & 72 & 84 & 78 & 97 & 116 & 114 & 101 & 106 & 97.9 & 80 & 101 & 67 & 79 & 78 & 92 & 100 & 84 & 96 & 114 & 89.1 \\
PIT & 115 & 106 & 81 & 100 & 82 & 116 & 94 & 90 & 121 & 82 & 98.7 & 110 & 73 & 94 & 109 & 56 & 82 & 82 & 100 & 94 & 86 & 88.6 \\
PHI & 65 & 85 & 81 & 74 & 118 & 72 & 90 & 79 & 83 & 104 & 85.1 & 80 & 77 & 91 & 86 & 108 & 46 & 59 & 78 & 97 & 100 & 82.2 \\
\bottomrule
\end{tabular}
\end{adjustbox}
\end{table*}

\begin{table}[htbp]
\small\sf\centering
\caption{Home and away TBR for selected team defense case studies, with corresponding Def$^{*}$ and DBS$^{*}$ values.\label{tab:vs_def}}
\begin{adjustbox}{max width=\columnwidth}
\begin{tabular}{lcccccc}
\toprule
\multirow{2}{*}{Team (Season)} & \multicolumn{2}{c}{TBR (Home)} & \multicolumn{2}{c}{TBR (Away)} & \multicolumn{2}{c}{Team Defense} \\
\cmidrule(lr){2-3} \cmidrule(lr){4-5} \cmidrule(lr){6-7}
 & Opp & Team & Opp & Team & Def$^{*}$ & DBS$^{*}$ \\
\midrule
OAK (2015) & -0.046 &  0.042 &  0.024 &  0.027 &  55 & 124 \\
BAL (2020) & -0.035 &  0.047 &  0.039 &  0.066 &  67 & 116 \\
SD (2016)  &  0.013 &  0.038 &  0.026 &  0.028 &  68 & 114 \\
\midrule
HOU (2016) &  0.051 &  0.019 &  0.044 &  0.019 & 134 &  72 \\
MIL (2020) &  0.049 & -0.016 &  0.015 & -0.027 & 124 &  70 \\
KC (2023)  & -0.023 & -0.065 & -0.006 & -0.045 & 120 &  69 \\
SF (2023)  & -0.038 & -0.062 &  0.007 & -0.045 & 120 &  70 \\
NYY (2024) & -0.035 & -0.072 & -0.036 & -0.059 & 128 &  82 \\
\bottomrule
\end{tabular}
\end{adjustbox}
\end{table}

\begin{table*}[htbp]
\small\sf\centering
\caption{Comparative View of OAA$^{*}$ and DBS$^{*}$ (2016--2024). Avg denotes the nine-year arithmetic average. Rows are ordered by the nine-year average of DBS$^{*}$, from strongest to weakest team defense.\label{tab:std_defense_comparison_OAA}}
\begin{adjustbox}{max width=\textwidth}
\begin{tabular}{lrrrrrrrrrr|rrrrrrrrrr}
\toprule
Team & \multicolumn{10}{c|}{OAA$^{*}$} & \multicolumn{10}{c}{DBS$^{*}$} \\
 & 2016 & 2017 & 2018 & 2019 & 2020 & 2021 & 2022 & 2023 & 2024 & Avg & 2016 & 2017 & 2018 & 2019 & 2020 & 2021 & 2022 & 2023 & 2024 & Avg \\
\midrule
TB & 74 & 111 & 100 & 104 & 97 & 125 & 101 & 101 & 105 & 102.0 & 109 & 114 & 115 & 85 & 112 & 139 & 118 & 132 & 121 & 116.1 \\
CHC & 148 & 118 & 122 & 130 & 115 & 110 & 78 & 117 & 121 & 117.7 & 136 & 127 & 135 & 123 & 94 & 91 & 99 & 128 & 108 & 115.7 \\
CLE & 117 & 126 & 92 & 129 & 113 & 116 & 119 & 112 & 110 & 114.9 & 111 & 121 & 60 & 115 & 115 & 124 & 144 & 127 & 117 & 114.9 \\
AZ & 92 & 99 & 125 & 125 & 121 & 88 & 141 & 129 & 132 & 116.9 & 107 & 122 & 139 & 138 & 93 & 95 & 118 & 125 & 80 & 113.0 \\
COL & 123 & 118 & 98 & 113 & 107 & 102 & 99 & 99 & 108 & 107.4 & 115 & 112 & 113 & 126 & 121 & 102 & 109 & 91 & 101 & 110.0 \\
LAD & 84 & 83 & 88 & 103 & 117 & 92 & 107 & 97 & 97 & 96.4 & 101 & 91 & 87 & 115 & 127 & 120 & 123 & 109 & 113 & 109.6 \\
MIL & 63 & 124 & 130 & 103 & 107 & 82 & 104 & 137 & 126 & 108.4 & 96 & 111 & 120 & 93 & 70 & 112 & 86 & 120 & 157 & 107.2 \\
TEX & 122 & 103 & 110 & 92 & 87 & 123 & 97 & 119 & 123 & 108.4 & 126 & 120 & 101 & 99 & 85 & 89 & 103 & 122 & 111 & 106.2 \\
BOS & 92 & 106 & 100 & 106 & 91 & 68 & 89 & 52 & 84 & 87.6 & 150 & 117 & 89 & 80 & 121 & 73 & 115 & 79 & 121 & 105.0 \\
HOU & 105 & 101 & 135 & 141 & 119 & 133 & 132 & 106 & 98 & 118.9 & 72 & 78 & 103 & 113 & 89 & 116 & 133 & 119 & 96 & 102.1 \\
KC & 127 & 115 & 89 & 85 & 121 & 119 & 100 & 130 & 132 & 113.1 & 113 & 88 & 82 & 104 & 112 & 134 & 90 & 69 & 125 & 101.9 \\
SD & 64 & 76 & 108 & 76 & 137 & 109 & 127 & 127 & 106 & 103.3 & 114 & 74 & 111 & 84 & 130 & 106 & 95 & 94 & 104 & 101.3 \\
TOR & 102 & 67 & 67 & 77 & 75 & 92 & 106 & 109 & 120 & 90.6 & 129 & 64 & 59 & 102 & 92 & 102 & 124 & 112 & 123 & 100.8 \\
ATL & 109 & 79 & 137 & 102 & 123 & 106 & 106 & 87 & 100 & 105.4 & 81 & 85 & 119 & 117 & 101 & 125 & 94 & 109 & 71 & 100.2 \\
SF & 116 & 89 & 97 & 113 & 101 & 122 & 69 & 117 & 94 & 102.0 & 121 & 88 & 126 & 112 & 116 & 112 & 68 & 70 & 78 & 99.0 \\
STL & 95 & 92 & 69 & 126 & 123 & 145 & 127 & 97 & 112 & 109.6 & 78 & 75 & 104 & 102 & 124 & 134 & 111 & 72 & 89 & 98.8 \\
WSH & 91 & 87 & 100 & 124 & 75 & 78 & 53 & 99 & 82 & 87.7 & 71 & 115 & 105 & 135 & 81 & 90 & 71 & 118 & 93 & 97.7 \\
CWS & 107 & 108 & 101 & 86 & 115 & 91 & 84 & 83 & 66 & 93.4 & 104 & 112 & 111 & 86 & 136 & 105 & 81 & 79 & 64 & 97.6 \\
MIN & 108 & 154 & 105 & 81 & 117 & 108 & 86 & 87 & 102 & 105.3 & 99 & 126 & 87 & 101 & 120 & 73 & 85 & 89 & 97 & 97.4 \\
BAL & 105 & 79 & 65 & 72 & 83 & 75 & 108 & 90 & 89 & 85.1 & 77 & 89 & 72 & 94 & 116 & 100 & 100 & 119 & 109 & 97.3 \\
SEA & 105 & 117 & 92 & 80 & 103 & 96 & 100 & 122 & 86 & 100.1 & 94 & 110 & 108 & 82 & 80 & 91 & 105 & 107 & 85 & 95.8 \\
OAK & 80 & 83 & 117 & 103 & 83 & 117 & 93 & 81 & 58 & 90.6 & 93 & 117 & 118 & 126 & 84 & 101 & 92 & 67 & 62 & 95.6 \\
LAA & 96 & 108 & 101 & 106 & 61 & 81 & 101 & 82 & 66 & 89.1 & 99 & 113 & 94 & 86 & 89 & 68 & 114 & 86 & 98 & 94.1 \\
DET & 97 & 89 & 83 & 85 & 115 & 87 & 113 & 91 & 120 & 97.8 & 73 & 81 & 72 & 67 & 96 & 107 & 124 & 119 & 97 & 92.9 \\
CIN & 85 & 110 & 115 & 123 & 81 & 68 & 82 & 67 & 87 & 90.9 & 96 & 61 & 114 & 106 & 88 & 77 & 81 & 90 & 110 & 91.4 \\
NYY & 111 & 116 & 70 & 67 & 83 & 81 & 122 & 104 & 108 & 95.8 & 86 & 125 & 95 & 67 & 100 & 78 & 98 & 88 & 82 & 91.0 \\
NYM & 60 & 57 & 89 & 85 & 93 & 121 & 107 & 87 & 106 & 89.4 & 101 & 67 & 79 & 78 & 92 & 100 & 84 & 96 & 114 & 90.1 \\
MIA & 121 & 93 & 122 & 91 & 75 & 93 & 104 & 69 & 66 & 92.7 & 98 & 113 & 88 & 100 & 88 & 94 & 57 & 71 & 87 & 88.4 \\
PIT & 100 & 87 & 105 & 77 & 103 & 92 & 81 & 103 & 87 & 92.8 & 73 & 94 & 109 & 56 & 82 & 82 & 100 & 94 & 86 & 86.2 \\
PHI & 97 & 102 & 71 & 98 & 63 & 81 & 67 & 100 & 108 & 87.4 & 77 & 91 & 86 & 108 & 46 & 59 & 78 & 97 & 100 & 82.4 \\
\bottomrule
\end{tabular}
\end{adjustbox}
\end{table*}

\begin{table}[htbp]
\small\sf\centering
\caption{Home and away TBR for selected team defense case studies, with corresponding OAA$^{*}$ and DBS$^{*}$ values.\label{tab:vs_oaa}}
\begin{adjustbox}{max width=\columnwidth}
\begin{tabular}{lcccccc}
\toprule
\multirow{2}{*}{Team (Season)} & \multicolumn{2}{c}{TBR (Home)} & \multicolumn{2}{c}{TBR (Away)} & \multicolumn{2}{c}{Team Defense} \\
\cmidrule(lr){2-3} \cmidrule(lr){4-5} \cmidrule(lr){6-7}
 & Opp & Team & Opp & Team & OAA$^{*}$ & DBS$^{*}$ \\
\midrule
KC (2023)  & -0.023 & -0.065 & -0.006 & -0.045 & 130 &  69 \\
AZ (2024)  & -0.031 & -0.015 & -0.006 &  0.001 & 132 &  80 \\
CIN (2017) &  0.115 &  0.090 &  0.073 &  0.046 & 110 &  61 \\
SF (2023)  & -0.038 & -0.062 &  0.007 & -0.045 & 117 &  70 \\
\midrule
SD (2016)  &  0.013 &  0.038 &  0.026 &  0.028 &  64 & 114 \\
BOS (2016) &  0.030 &  0.074 & -0.035 &  0.042 &  92 & 150 \\
BOS (2024) & -0.041 & -0.001 & -0.039 & -0.023 &  84 & 121 \\
OAK (2017) &  0.011 &  0.044 &  0.029 & -0.001 &  83 & 117 \\
\bottomrule
\end{tabular}
\end{adjustbox}
\end{table}

\begin{table}[htbp]
\small\sf\centering
\caption{Correlation of DBS$^{*}$ with Def$^{*}$ and OAA$^{*}$, 2015--2024.\label{tab:defense_correlation}}
\begin{adjustbox}{max width=\columnwidth}
\begin{tabular}{lcc}
\toprule
Season & DBS$^{*}$ vs.\ Def$^{*}$ & DBS$^{*}$ vs.\ OAA$^{*}$ \\ 
\midrule
2015 & 0.3626 & -- \\ 
2016 & 0.2700 & 0.2179 \\ 
2017 & 0.5247 & 0.6167 \\ 
2018 & 0.4099 & 0.6287 \\ 
2019 & 0.5513 & 0.6668 \\ 
2020 & 0.4406 & 0.5508 \\ 
2021 & 0.4511 & 0.6256 \\ 
2022 & 0.5316 & 0.5882 \\ 
2023 & 0.3996 & 0.3982 \\ 
2024 & 0.4679 & 0.5013 \\ 
\midrule
Mean & 0.4409 & 0.5327 \\ 
\bottomrule
\end{tabular}
\end{adjustbox}
\end{table}

The uncertainty estimates quantify the precision of the season-specific defensive effects. Across the 300 team-season estimates, the average half-width of the 95\% confidence intervals for DBS$^{*}$ is 30.84 index points (SD = 3.74). These intervals are wider than those for TBR-PF$^{*}$ on the standardized scale because the estimated defensive effects are less dispersed across teams relative to their standard errors. This does not make DBS$^{*}$ uninformative; rather, the intervals indicate how much evidence the available data provide that a team's season-specific defensive effect differs from the league average. The full confidence intervals and the distribution of their half-widths are reported in Appendix~\ref{app:defense_ci}.

We first compare DBS$^{*}$ with Def$^{*}$. Table~\ref{tab:std_defense_comparison} reports the corresponding standardized indices defined in equation~\eqref{eq:index}, which place Def$^{*}$ and DBS$^{*}$ on a comparable scale. To further investigate their differences, we identify eight cases with the largest discrepancies between the two measures and report them in Table~\ref{tab:vs_def}. For ease of interpretation, we group the cases into two categories. The first three correspond to teams for which Def$^{*}$ is below 100 while DBS$^{*}$ is above 100, and the remaining five correspond to the opposite pattern.

For the first group, we observe that, in all three cases, the TBR of the opposing teams is lower than that of the listed team in both home and away settings. Since TBR is already defined conditional on batted-ball quality, and here further separated by home and away contexts, lower TBR values indicate better defensive outcomes. Therefore, these patterns suggest that the teams exhibit stronger defensive performance relative to their opponents. Our estimates, which place these teams above the league average, are consistent with the observed outcome patterns.

Similarly, for the second group, the opposing teams consistently exhibit higher TBR in both home and away games, indicating that the teams have weaker defensive performance. This is again consistent with our estimates, which assign these teams defensive effects below the league average. These results show that the estimates from our proposed model are broadly consistent with observed batted-ball outcomes when compared with Def.

We next compare DBS$^{*}$ with OAA$^{*}$. As before, we standardize both measures using equation~\eqref{eq:index} and report them side by side in Table~\ref{tab:std_defense_comparison_OAA}. 

To further examine their differences, Table~\ref{tab:vs_oaa} lists the cases with the largest discrepancies between the two measures. In these cases, the patterns of TBR are mostly consistent with our estimates. The two exceptions are discussed below.

 For OAK (2017), the results are mixed. At home, the TBR values suggest that OAK exhibits better defensive outcomes than its opponents, while in away games the opposite pattern is observed. The difference is slightly larger in home games than in away games, suggesting that the evidence slightly favors our estimate, although the overall results do not clearly support either measure.

A more notable discrepancy arises for AZ (2024). The TBR values suggest that AZ performs better than its opponents in both home and away games, whereas our estimate places the team below the league average. This indicates that our estimate is less aligned with the observed outcomes in this case. However, the interpretation changes when the league-wide baseline is taken into account. In 2024, the average TBR across all games is -0.0364, which is below the overall average across the 2015--2024 period. At Chase Field (AZ), the opponents’ TBR is approximately -0.031, which is lower than that of their opponents but still higher than the league-wide average. From this perspective, AZ’s defensive performance can still be viewed as below average, while the apparent advantage relative to their opponents may partly reflect even weaker defensive performance by those opponents.

Overall, the comparison with OAA is less clear-cut than the comparison with Def. However, in the majority of cases, the patterns of TBR are more consistent with the estimates from our proposed model than with OAA.

We next compare which existing metric is more closely aligned with our estimates by examining the correlations over time. Table~\ref{tab:defense_correlation} reports the correlations of DBS$^{*}$ with Def$^{*}$ and OAA$^{*}$ across seasons. OAA is not available in 2015, as the data required to compute OAA were not fully collected in that year. As expected, OAA$^{*}$, which is conceptually closer to our framework, exhibits consistently higher correlations with DBS$^{*}$ than Def$^{*}$. However, the magnitude of the correlations remains moderate, with the highest value reaching 0.6668.

\subsection{Changes in the League-Wide Baseline Over Time}

In this subsection, we examine the estimated intercept across seasons. 
Since both ballpark effects and team defensive effects are centered to have mean zero, the intercept reflects the league-wide average level of TBR in each season, relative to the overall average across all seasons.

Changes in the intercept therefore capture league-wide shifts that are not attributed to particular ballparks or defensive teams. These shifts may reflect changes in overall defensive performance, the physical properties of the baseball, or other league-wide conditions. Lower values of the intercept indicate that, on average, batted balls result in fewer bases than expected relative to the common baseline.

\begin{figure}[htbp]
    \centering
    \includegraphics[width=\columnwidth]{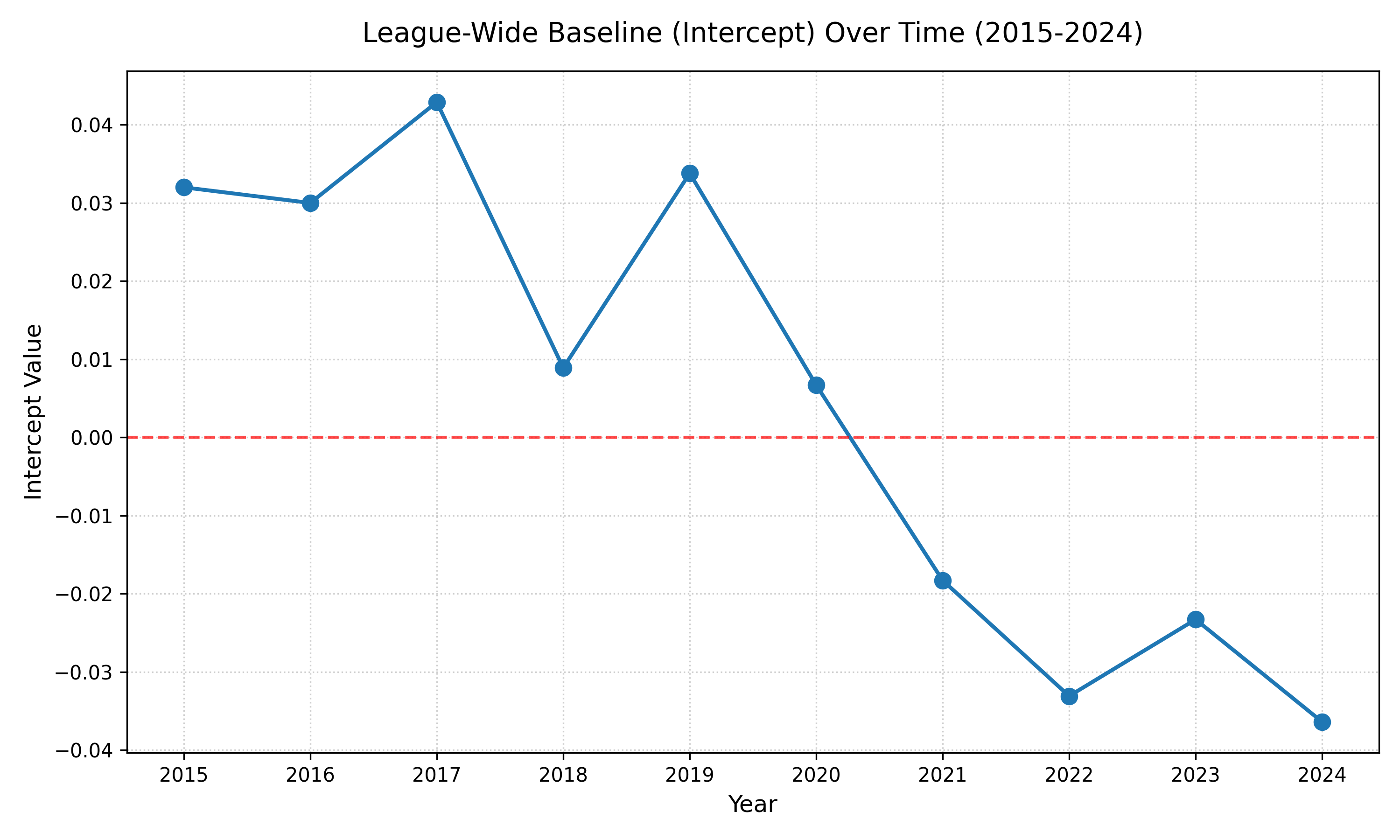}
    \caption{Yearly Comparison of League-Wide Intercept Over Time (2015--2024).}
    \label{fig:intercept_trends}
\end{figure}

As shown in Figure~\ref{fig:intercept_trends}, the intercept exhibits a clear downward trend over time. From 2015 to 2019, the intercept is positive, indicating that outcomes are above the baseline. Starting in the early 2020s, the intercept becomes negative, reaching $-0.0331$ in 2022, which indicates a shift in league-wide outcomes relative to the baseline.

It is worth noting that this period coincides with well-documented changes in the physical properties of the baseball. In particular, the seasons from 2017 to 2019 are often referred to as the “juiced ball” era, during which reduced aerodynamic drag led to increased carry of batted balls. In contrast, subsequent seasons saw efforts by MLB to reduce ball liveliness, including adjustments to ball construction and the introduction of humidors across ballparks. These changes are consistent with the elevated intercept values observed during 2017–2019 and the subsequent decline in later seasons.

The observed trend is therefore consistent with multiple developments over this period. On one hand, the increased use of data-driven defensive positioning may have contributed to suppressing offensive outcomes. On the other hand, changes in the physical properties of the baseball may also have played an important role. In addition, MLB introduced restrictions on extreme infield shifts prior to the 2023 season, and the upward movement of the intercept in 2023 aligns with this change. Taken together, these patterns suggest that the estimated intercept captures league-wide trends that are broadly consistent with developments in the game.

\section{Conclusion and Discussion}

In this paper, we propose a simple statistical framework for jointly estimating ballpark effects and team-level defense using total bases residuals (TBR). By controlling for contact quality through exit velocity and launch angle, the model separates the effects of ballpark environment and team defense within a unified regression framework.

Empirically, our results provide several insights. For ballpark effects, our estimates are broadly consistent with observed home–away patterns, and in cases where they differ from official MLB park factors, they are more consistent with observed home–away patterns for both the focal team and their opponents. For team defense, our estimates show moderate agreement with existing metrics and tend to be more consistent with batted-ball outcomes. These findings suggest that the proposed framework provides a meaningful representation of the underlying factors affecting batted-ball outcomes.

The estimated league-wide intercept provides an additional perspective on changes in the game over time. The intercept exhibits a clear downward trend from 2015 to 2022, suggesting that, on average, batted-ball outcomes became less favorable to offense relative to the baseline. This pattern is broadly consistent with the increased use of data-driven defensive positioning and possible changes in the physical properties of the baseball. The upward movement observed in 2023 aligns with the introduction of restrictions on extreme infield shifts.

We also introduce a general standardization method based on the z-score transformation, which provides a more interpretable 100-based scale for comparing effect magnitudes. Applied to our estimates, this transformation yields TBR-PF$^{*}$ and DBS$^{*}$; the same transformation can also be applied to other metrics measured across comparable units, as illustrated by MLB PF$^{*}$, Def$^{*}$, and OAA$^{*}$. Unlike traditional ratio-based measures, this standardized scale directly reflects relative standing within the relevant distribution and allows meaningful comparisons of relative standing across different metrics and across teams, ballparks, and seasons.

A further advantage of the regression framework is that it provides explicit uncertainty quantification for both ballpark and defensive effects. On the standardized scale, the average 95\% confidence-interval half-width is 19.97 points for TBR-PF$^{*}$ and 30.84 points for DBS$^{*}$. For ballpark effects, this result shows that part of the apparent year-to-year instability of single-season estimates may reflect sampling variability. The point estimates nevertheless remain useful as season-specific assessments, while the confidence intervals describe the precision attainable from the available data.

Sampling variability is not the only source of year-to-year variation in park effects. Although ballpark characteristics such as dimensions remain fixed over time, our results suggest that estimated park effects may still vary across seasons. One possible explanation is that many ballparks are not symmetric and differ in outfield size, so their effects depend on the distribution of batted-ball outcomes. For example, differences between left and right field dimensions may lead to different outcomes depending on where the ball is hit. Similarly, smaller outfields may increase the probability of home runs while reducing the space for balls to fall in play. As a result, a single park factor may reflect a combination of these effects, which can vary from year to year.

From a practical perspective, the proposed framework provides a transparent tool for evaluating how much of a team’s observed performance is attributable to the ballpark versus the defense they face. Because defensive performance is inherently difficult to measure and no single metric is universally accepted, our approach offers a simple and coherent way to estimate team-level defense within the same framework. The resulting measures can be used to compare team-level defensive performance across seasons and to interpret offensive and defensive outcomes across different ballpark environments.

Several limitations should be noted. The current model does not incorporate the direction of batted balls, which may limit its ability to capture asymmetric park effects. In addition, total bases are used as a proxy for offensive outcomes and are not directly proportional to run value. The model also assumes an additive structure between ballpark and defensive effects, which may not fully capture potential interactions. Addressing these limitations and extending the framework are important directions for future work.

\bibliographystyle{SageH}
\bibliography{bibliography.bib}

%\begin{appendices}
\appendix
\onecolumn
%\renewcommand{\thetable}{\thesection}
% --- Appendix A ---
\section{Comparison of Offensive Performance for the Minnesota Twins and Opponents in 2021} \label{app:min_statcast_2021}
\setcounter{table}{0}
\renewcommand{\thetable}{A.\arabic{table}}

\begin{table*}[htbp]
    \centering
    \caption{2021 Minnesota Twins Statcast Batting Summary}
    \label{tab:min_statcast_2021}
    \begin{tabular}{lrrrrrrrr}
        \toprule
        \textbf{Team (Location)} & \textbf{G} & \textbf{PA} & \textbf{HIP} & \textbf{SO/PA} & \textbf{SLG}& \textbf{TB/HIP}  & \textbf{HR/HIP} & \textbf{R/G} \\
        \midrule
        Minnesota (Target Field) & 81 & 3004 & 2007 & 0.224 &0.420& 0.554  & 0.056 & 4.51 \\
        Minnesota (other ballparks) & 81 & 3074 & 2068 & 0.238 &0.427& 0.574  & 0.056 & 4.49 \\
        Opponents (Target Field) & 81 & 3111 & 2198 & 0.211 &0.446& 0.570  & 0.055 & 5.05 \\
        Opponents (other ballparks) & 81 & 2969 & 2016 & 0.222 &0.454& 0.596  & 0.059 & 5.25 \\
        \midrule
        All teams (Target Field) & 162 & 6115 & 4205 & 0.217 & 0.433 &0.562  & 0.055 & 4.78 \\
        All teams (all AL ballparks) & 2430 & 90886 & 61257 & 0.230 & 0.414 &0.549 & 0.050 & 4.58 \\
        \bottomrule
    \end{tabular}
\end{table*}

We first clarify the notation used in Table~\ref{tab:min_statcast_2021}. HIP denotes the number of balls hit into play, including all batted-ball outcomes such as home runs. Based on this definition, TB/HIP represents total bases per ball hit into play, and HR/HIP represents home runs per ball hit into play.

We begin with standard measures. Based on slugging percentage (SLG), both Minnesota and their opponents show lower offensive production in Target Field than in their games at other ballparks. This suggests that, when comparing the same teams across locations, Target Field appears to suppress offensive outcomes.

However, a different pattern emerges when we look at overall comparisons. When all teams are pooled together, offensive production in Target Field appears comparable to, or slightly higher than, that in other American League ballparks. This apparent discrepancy reflects the effect of aggregation across teams.

To better isolate the effect of contact, we consider TB/HIP, which measures total bases per ball hit into play. Under this measure, the difference becomes more pronounced: both Minnesota and their opponents generate fewer total bases per batted ball in Target Field than elsewhere. This indicates that, conditional on contact, offensive outcomes are clearly lower in Target Field.

One reason for this difference is that the strikeout rate in Target Field is relatively low. Because SLG is defined per at-bat, it is affected by the frequency of strikeouts. A lower strikeout rate can increase the number of balls put into play, which in turn can make overall SLG appear closer to that of other ballparks, even if the quality of contact outcomes is lower. In contrast, TB/HIP focuses only on batted-ball outcomes and is therefore less affected by variation in strikeout rates.

\newpage
% --- Appendix B ---
\section{Original MLB Official Park Factors} \label{app:mlb_original}

\setcounter{table}{0}
\renewcommand{\thetable}{B.\arabic{table}}

Table \ref{tab:original_mlb_appendix} lists the raw, non-standardized park factors as published by Major League Baseball. These values represent the baseline against which our standardized comparisons were conducted.

\begin{table*}[htbp]
\small\sf\centering
\caption{Original MLB Official Park Factors (Non-Standardized), 2015--2024\label{tab:original_mlb_appendix}}
\begin{adjustbox}{max width=\textwidth}
\begin{tabular}{lrrrrrrrrrr}
\toprule
Team & 2015 & 2016 & 2017 & 2018 & 2019 & 2020 & 2021 & 2022 & 2023 & 2024 \\
\midrule
ATL & 97 & 99 & 101 & 98 & 104 & 101 & 100 & 99 & 103 & 99 \\
AZ & 105 & 110 & 105 & 100 & 99 & 102 & 104 & 98 & 99 & 106 \\
BAL & 104 & 101 & 103 & 102 & 105 & 94 & 108 & 99 & 96 & 102 \\
BOS & 107 & 109 & 100 & 103 & 103 & 108 & 107 & 108 & 108 & 102 \\
CHC & 100 & 94 & 103 & 102 & 99 & 96 & 103 & 99 & 101 & 91 \\
CIN & 102 & 103 & 102 & 105 & 99 & 105 & 110 & 107 & 102 & 105 \\
CLE & 104 & 105 & 96 & 103 & 97 & 100 & 100 & 96 & 94 & 102 \\
COL & 117 & 115 & 112 & 115 & 118 & 107 & 108 & 114 & 113 & 110 \\
CWS & 97 & 99 & 101 & 98 & 99 & 102 & 101 & 101 & 99 & 99 \\
DET & 100 & 103 & 106 & 97 & 101 & 102 & 96 & 97 & 99 & 97 \\
HOU & 98 & 93 & 98 & 95 & 102 & 92 & 100 & 99 & 100 & 103 \\
KC & 99 & 99 & 97 & 101 & 100 & 102 & 102 & 105 & 106 & 102 \\
LAA & 94 & 97 & 95 & 97 & 100 & 108 & 101 & 100 & 100 & 101 \\
LAD & 95 & 94 & 95 & 97 & 97 & 102 & 99 & 101 & 99 & 100 \\
MIA & 96 & 94 & 98 & 92 & 96 & 98 & 94 & 100 & 101 & 104 \\
MIL & 103 & 100 & 101 & 99 & 99 & 102 & 96 & 98 & 98 & 96 \\
MIN & 102 & 104 & 104 & 102 & 100 & 94 & 101 & 100 & 100 & 105 \\
NYM & 93 & 96 & 96 & 91 & 97 & 101 & 96 & 96 & 97 & 99 \\
NYY & 101 & 99 & 99 & 104 & 97 & 101 & 97 & 98 & 99 & 103 \\
OAK & 96 & 92 & 103 & 95 & 96 & 94 & 94 & 97 & 96 & 98 \\
PHI & 103 & 97 & 102 & 98 & 103 & 103 & 101 & 103 & 99 & 100 \\
PIT & 97 & 104 & 98 & 100 & 102 & 98 & 102 & 101 & 98 & 102 \\
SD & 99 & 98 & 94 & 100 & 95 & 99 & 97 & 92 & 96 & 99 \\
SEA & 97 & 99 & 97 & 96 & 96 & 95 & 92 & 93 & 93 & 89 \\
SF & 94 & 100 & 93 & 98 & 91 & 103 & 98 & 100 & 94 & 96 \\
STL & 98 & 98 & 98 & 97 & 95 & 100 & 93 & 98 & 103 & 98 \\
TB & 95 & 95 & 95 & 95 & 93 & 95 & 94 & 96 & 98 & 96 \\
TEX & 107 & 106 & 107 & 111 & 107 & 100 & 97 & 102 & 106 & 95 \\
TOR & 99 & 101 & 99 & 102 & 100 & 101 & 95 & 103 & 97 & 100 \\
WSH & 100 & 96 & 102 & 106 & 105 & 97 & 104 & 102 & 103 & 99 \\
\bottomrule
\end{tabular}
\end{adjustbox}
\end{table*}

\newpage
% --- Appendix C ---
\section{Original MLB Official Defense Factors} \label{app:mlb_defense_official}

\setcounter{table}{0}
\renewcommand{\thetable}{C.\arabic{table}}

Table \ref{tab:defense_factor_appendix} presents the Defensive Runs Above Average (Def) metrics published by FanGraphs for the period 2015--2024. Positive values indicate a defensive contribution above the league average, whereas negative values signify below-average performance.

\begin{table*}[htbp]
\small\sf\centering
\caption{Defensive Runs Above Average (Def), 2015--2024\label{tab:defense_factor_appendix}}
\begin{adjustbox}{max width=\textwidth}
\begin{tabular}{lrrrrrrrrrr}
\toprule
Team & 2015 & 2016 & 2017 & 2018 & 2019 & 2020 & 2021 & 2022 & 2023 & 2024 \\
\midrule
ARI & -0.4 & -21.4 & -3.2 & 51.3 & 41.5 & 10.6 & -17.2 & 31.3 & 17.8 & 21.2 \\
ATL & -10.7 & -5.0 & -19.7 & 30.2 & 9.2 & 11.5 & 2.4 & 3.8 & 11.7 & 1.4 \\
BAL & 8.9 & 4.3 & -21.2 & -43.8 & -65.6 & -16.8 & -52.8 & 14.4 & 0.0 & -13.4 \\
BOS & -13.0 & 4.2 & 32.6 & 38.6 & 28.8 & 0.0 & -12.6 & -9.2 & -50.6 & -21.6 \\
CHC & 25.6 & 51.7 & 6.9 & 17.1 & 17.4 & 14.3 & 5.0 & -20.0 & 10.5 & 12.9 \\
CIN & -17.7 & -28.5 & 7.6 & -11.1 & 29.3 & -4.6 & -32.2 & -23.9 & -42.9 & -19.8 \\
CLE & 13.0 & 22.7 & 46.6 & 3.3 & 57.3 & 7.6 & 17.7 & 22.3 & 6.8 & 33.8 \\
COL & -36.4 & 22.1 & 4.1 & 2.5 & -0.6 & -1.9 & 10.0 & -13.7 & -0.3 & -3.7 \\
CWS & -6.0 & -13.3 & 9.3 & -27.6 & -36.4 & 15.1 & -33.2 & -11.3 & -23.8 & -56.2 \\
DET & -16.6 & -24.9 & -34.7 & -27.0 & -30.5 & -0.4 & -27.7 & -0.6 & 2.2 & 32.4 \\
HOU & -4.3 & 36.6 & 13.1 & 50.7 & 34.8 & 1.7 & 32.7 & 34.8 & -3.1 & 2.1 \\
KC & 29.0 & 12.7 & 0.6 & -20.6 & -11.2 & 8.8 & 11.1 & -25.2 & 23.1 & 35.4 \\
LAA & 17.8 & -0.4 & 42.9 & 15.4 & -5.8 & -18.9 & -21.8 & -16.8 & -29.5 & -36.4 \\
LAD & 22.8 & 17.5 & 15.1 & 13.9 & 23.6 & 4.9 & -2.6 & 11.9 & 1.1 & -12.0 \\
MIA & -5.0 & 13.0 & -4.9 & 28.4 & -17.8 & -13.4 & -1.4 & 1.4 & -27.9 & -17.1 \\
MIL & -10.6 & -16.1 & 23.5 & 46.0 & 20.6 & 12.5 & -3.0 & 11.5 & 50.4 & 29.1 \\
MIN & 13.3 & -23.0 & 58.1 & -1.4 & -10.7 & 6.4 & 15.5 & -4.3 & -5.5 & 6.4 \\
NYM & 13.3 & -2.3 & -35.0 & -21.3 & -31.4 & -1.5 & 19.9 & 16.7 & 1.4 & 7.3 \\
NYY & -20.1 & 15.3 & 19.1 & -10.7 & -28.0 & -5.5 & -26.1 & 59.3 & 17.3 & 34.8 \\
OAK & -45.3 & -46.5 & -32.4 & 7.1 & -6.5 & -3.8 & 24.3 & -6.1 & -34.4 & -49.5 \\
PHI & -34.4 & -16.1 & -23.8 & -34.0 & 24.8 & -14.6 & -12.2 & -25.4 & -20.0 & 4.4 \\
PIT & 15.5 & 6.9 & -23.5 & 0.2 & -25.8 & 8.6 & -7.0 & -11.9 & 25.4 & -22.2 \\
SD & -0.2 & -35.2 & -13.6 & 6.0 & -7.5 & 13.5 & 6.9 & 1.7 & 16.2 & -8.5 \\
SEA & 0.3 & -15.3 & 21.3 & -6.2 & -37.1 & -4.6 & -24.8 & -0.3 & 20.0 & -2.7 \\
SF & 26.6 & 24.6 & -22.5 & -17.7 & 17.3 & -0.8 & 28.9 & -44.1 & 23.1 & 8.4 \\
STL & 18.4 & -7.5 & -9.0 & -36.1 & 11.8 & 10.4 & 41.1 & 24.4 & -19.9 & -3.3 \\
TB & 20.5 & -4.9 & 10.1 & -2.5 & 16.2 & 4.4 & 27.4 & 1.8 & -8.0 & 3.5 \\
TEX & 13.4 & 11.0 & -9.5 & -19.1 & -27.6 & -15.2 & 45.5 & 2.4 & 31.1 & 19.2 \\
TOR & 21.7 & 11.3 & -34.6 & -20.1 & -3.4 & -10.8 & 8.7 & 22.1 & 23.0 & 36.9 \\
WSH & -23.6 & 3.7 & -23.7 & -8.5 & 10.2 & -14.9 & -21.4 & -57.6 & -22.3 & -29.6 \\
\bottomrule
\end{tabular}
\end{adjustbox}
\end{table*}

\newpage

% --- Appendix D ---
\section{Home and Away Defense Residual Analysis} \label{app:home_away_residuals}

\setcounter{table}{0}
\renewcommand{\thetable}{D.\arabic{table}}

Table \ref{tab:home_away_appendix} presents a comprehensive Total Bases Residual (TBR) analysis for all 30 MLB teams spanning the 2015--2024 seasons. To precisely capture how stadium environments dynamically interact with both home and visiting lineups, the data is structured into four distinct situational metrics. Specifically, \textit{Team TBR (home)} and \textit{Team TBR (away)} represent the focal team's own offensive TBR when batting at their home stadium and on the road, respectively. Conversely, \textit{Opp TBR (home)} and \textit{Opp TBR (away)} denote the TBR generated by opposing batters when the focal team is playing defense at their home ballpark and at away venues, respectively. By evaluating the discrepancy between actual total bases and their expected values across these four dimensions, the analysis provides a granular view of venue-specific impacts on both offensive and defensive outcomes.

\begingroup
\setlength{\tabcolsep}{4pt} % 縮小欄位間距
\footnotesize               % 縮小字體
\begin{longtable}{lrrrrrrrrrr} % 拿掉直線 |
\caption{\normalsize Home and Away Total Bases Residual (TBR) Analysis by Team (2015--2024)} \label{tab:home_away_appendix} \\
\toprule
Metric & 2015 & 2016 & 2017 & 2018 & 2019 & 2020 & 2021 & 2022 & 2023 & 2024 \\
\midrule
\endfirsthead
\multicolumn{11}{c}{{\bfseries \tablename\ \thetable{} -- continued from previous page}} \\
\toprule
Metric & 2015 & 2016 & 2017 & 2018 & 2019 & 2020 & 2021 & 2022 & 2023 & 2024 \\
\midrule
\endhead
\midrule
\multicolumn{11}{r}{{Continued on next page}} \\
\endfoot
\bottomrule
\noalign{\vspace{0.8em}} 
\label{tab:home_away_appendix}
\endlastfoot

\textbf{ATL} & ~ & ~ & ~ & ~ & ~ & ~ & ~ & ~ & ~ & ~ \\ \midrule
Opp TBR (home) & -0.013 & -0.007 & 0.049 & -0.040 & 0.004 & -0.006 & -0.045 & -0.055 & -0.029 & -0.022 \\ 
Opp TBR (away) & 0.062 & 0.035 & 0.045 & 0.008 & 0.020 & 0.023 & -0.036 & -0.028 & -0.033 & -0.029 \\ 
Team TBR (home) & -0.004 & -0.022 & 0.027 & 0.012 & 0.024 & 0.024 & -0.016 & -0.041 & -0.021 & -0.072 \\ 
Team TBR (away) & -0.001 & 0.018 & 0.038 & 0.039 & -0.002 & 0.005 & -0.056 & -0.044 & -0.037 & -0.058 \\ \midrule
\textbf{AZ} & ~ & ~ & ~ & ~ & ~ & ~ & ~ & ~ & ~ & ~ \\ \midrule
Opp TBR (home) & 0.030 & 0.043 & 0.009 & -0.039 & -0.008 & 0.000 & 0.026 & -0.069 & -0.024 & -0.031 \\ 
Opp TBR (away) & 0.014 & 0.026 & 0.051 & -0.015 & 0.005 & 0.027 & -0.006 & -0.025 & -0.031 & -0.006 \\ 
Team TBR (home) & 0.031 & 0.054 & 0.078 & 0.003 & 0.030 & 0.002 & 0.032 & -0.008 & 0.026 & -0.015 \\ 
Team TBR (away) & 0.041 & 0.015 & 0.043 & -0.001 & 0.059 & -0.021 & -0.018 & -0.010 & -0.029 & 0.001 \\ \midrule
\textbf{BAL} & ~ & ~ & ~ & ~ & ~ & ~ & ~ & ~ & ~ & ~ \\ \midrule
Opp TBR (home) & 0.050 & 0.038 & 0.030 & 0.044 & 0.066 & -0.035 & -0.007 & -0.057 & -0.080 & -0.053 \\ 
Opp TBR (away) & 0.042 & 0.032 & 0.058 & 0.027 & 0.021 & 0.039 & -0.019 & -0.031 & -0.028 & -0.045 \\ 
Team TBR (home) & 0.047 & 0.007 & 0.034 & 0.022 & 0.042 & 0.047 & -0.002 & -0.067 & -0.042 & -0.052 \\ 
Team TBR (away) & -0.013 & -0.010 & 0.019 & -0.017 & 0.019 & 0.066 & -0.015 & -0.028 & 0.004 & -0.009 \\ \midrule
\textbf{BOS} & ~ & ~ & ~ & ~ & ~ & ~ & ~ & ~ & ~ & ~ \\ \midrule
Opp TBR (home) & 0.030 & 0.030 & 0.017 & -0.004 & 0.022 & 0.069 & 0.010 & -0.016 & 0.011 & -0.041 \\ 
Opp TBR (away) & 0.034 & -0.035 & 0.026 & 0.013 & 0.061 & -0.015 & 0.007 & -0.055 & 0.001 & -0.039 \\ 
Team TBR (home) & 0.065 & 0.074 & 0.036 & -0.002 & 0.034 & 0.084 & -0.009 & -0.022 & 0.007 & -0.001 \\ 
Team TBR (away) & 0.009 & 0.042 & 0.023 & -0.009 & 0.018 & 0.035 & -0.048 & -0.049 & -0.032 & -0.023 \\ \midrule
\textbf{CHC} & ~ & ~ & ~ & ~ & ~ & ~ & ~ & ~ & ~ & ~ \\ \midrule
Opp TBR (home) & 0.028 & -0.026 & 0.040 & -0.012 & -0.002 & -0.035 & 0.002 & -0.031 & -0.029 & -0.070 \\ 
Opp TBR (away) & 0.022 & -0.020 & 0.004 & -0.034 & 0.023 & 0.023 & 0.006 & -0.018 & -0.048 & -0.034 \\ 
Team TBR (home) & 0.021 & 0.021 & 0.036 & 0.009 & 0.042 & -0.022 & 0.016 & 0.003 & -0.004 & -0.055 \\ 
Team TBR (away) & 0.028 & 0.041 & 0.053 & 0.006 & 0.042 & 0.021 & 0.002 & -0.027 & -0.016 & -0.034 \\ \midrule
\textbf{CIN} & ~ & ~ & ~ & ~ & ~ & ~ & ~ & ~ & ~ & ~ \\ \midrule
Opp TBR (home) & 0.056 & 0.055 & 0.115 & 0.069 & 0.067 & 0.076 & 0.059 & 0.051 & 0.036 & 0.020 \\ 
Opp TBR (away) & 0.001 & 0.029 & 0.073 & -0.024 & 0.018 & -0.004 & 0.002 & -0.025 & -0.014 & -0.053 \\ 
Team TBR (home) & 0.045 & 0.065 & 0.090 & 0.048 & 0.065 & 0.058 & 0.030 & 0.019 & 0.038 & 0.016 \\ 
Team TBR (away) & -0.002 & 0.013 & 0.046 & -0.019 & 0.023 & -0.083 & -0.047 & -0.047 & 0.010 & -0.039 \\ \midrule
\textbf{CLE} & ~ & ~ & ~ & ~ & ~ & ~ & ~ & ~ & ~ & ~ \\ \midrule
Opp TBR (home) & 0.014 & 0.030 & 0.033 & 0.016 & 0.052 & 0.019 & -0.025 & -0.066 & -0.072 & -0.042 \\ 
Opp TBR (away) & -0.003 & 0.001 & 0.010 & 0.034 & 0.001 & -0.043 & -0.046 & -0.078 & -0.040 & -0.029 \\ 
Team TBR (home) & 0.012 & 0.049 & 0.035 & -0.004 & 0.057 & 0.010 & -0.001 & -0.024 & -0.027 & 0.007 \\ 
Team TBR (away) & 0.004 & 0.015 & 0.045 & -0.004 & 0.024 & -0.056 & -0.022 & 0.010 & -0.025 & -0.010 \\ \midrule
\textbf{COL} & ~ & ~ & ~ & ~ & ~ & ~ & ~ & ~ & ~ & ~ \\ \midrule
Opp TBR (home) & 0.095 & 0.081 & 0.119 & 0.078 & 0.092 & 0.046 & 0.018 & 0.015 & 0.021 & 0.009 \\ 
Opp TBR (away) & 0.024 & -0.002 & 0.032 & 0.006 & 0.006 & -0.022 & 0.006 & -0.041 & -0.008 & -0.030 \\ 
Team TBR (home) & 0.107 & 0.109 & 0.148 & 0.097 & 0.116 & 0.074 & 0.060 & 0.039 & 0.031 & 0.025 \\ 
Team TBR (away) & 0.016 & 0.020 & 0.039 & -0.005 & 0.047 & -0.003 & -0.038 & -0.034 & -0.034 & -0.058 \\ \midrule
\textbf{CWS} & ~ & ~ & ~ & ~ & ~ & ~ & ~ & ~ & ~ & ~ \\ \midrule
Opp TBR (home) & 0.038 & 0.039 & 0.037 & -0.012 & 0.029 & -0.023 & -0.006 & -0.015 & -0.019 & -0.019 \\ 
Opp TBR (away) & 0.048 & 0.023 & 0.042 & -0.007 & 0.034 & -0.035 & -0.029 & -0.031 & -0.002 & -0.020 \\ 
Team TBR (home) & 0.028 & 0.053 & 0.081 & 0.038 & 0.013 & 0.035 & 0.003 & -0.059 & -0.024 & -0.061 \\ 
Team TBR (away) & 0.020 & 0.025 & 0.016 & -0.004 & 0.027 & 0.016 & -0.022 & -0.047 & -0.025 & -0.065 \\ \midrule
\textbf{DET} & ~ & ~ & ~ & ~ & ~ & ~ & ~ & ~ & ~ & ~ \\ \midrule
Opp TBR (home) & 0.023 & 0.020 & 0.003 & -0.025 & 0.029 & -0.030 & -0.076 & -0.079 & -0.045 & -0.049 \\ 
Opp TBR (away) & 0.066 & 0.033 & 0.064 & 0.045 & 0.051 & 0.033 & -0.010 & -0.067 & -0.060 & -0.032 \\ 
Team TBR (home) & 0.000 & -0.028 & 0.002 & -0.018 & -0.005 & -0.007 & -0.036 & -0.072 & -0.062 & -0.041 \\ 
Team TBR (away) & 0.012 & -0.010 & -0.004 & -0.013 & 0.013 & 0.014 & -0.010 & -0.047 & -0.050 & -0.042 \\ \midrule
\textbf{HOU} & ~ & ~ & ~ & ~ & ~ & ~ & ~ & ~ & ~ & ~ \\ \midrule
Opp TBR (home) & 0.035 & 0.051 & 0.057 & 0.037 & 0.067 & -0.019 & -0.017 & -0.050 & -0.017 & 0.000 \\ 
Opp TBR (away) & 0.016 & 0.044 & 0.074 & -0.010 & 0.024 & 0.026 & -0.051 & -0.066 & -0.065 & -0.054 \\ 
Team TBR (home) & 0.116 & 0.019 & 0.078 & 0.036 & 0.122 & 0.024 & -0.009 & -0.016 & -0.045 & -0.032 \\ 
Team TBR (away) & 0.002 & 0.019 & 0.050 & 0.025 & 0.029 & -0.009 & -0.026 & -0.052 & -0.012 & -0.039 \\ \midrule
\textbf{KC} & ~ & ~ & ~ & ~ & ~ & ~ & ~ & ~ & ~ & ~ \\ \midrule
Opp TBR (home) & -0.026 & -0.037 & 0.017 & -0.024 & -0.009 & -0.053 & -0.061 & -0.063 & -0.023 & -0.059 \\ 
Opp TBR (away) & 0.015 & 0.034 & 0.047 & 0.017 & 0.026 & 0.021 & -0.066 & -0.013 & -0.006 & -0.058 \\ 
Team TBR (home) & 0.011 & 0.017 & 0.014 & -0.024 & -0.001 & -0.001 & -0.044 & -0.056 & -0.065 & -0.051 \\ 
Team TBR (away) & 0.016 & 0.015 & 0.012 & -0.018 & 0.033 & 0.009 & -0.035 & -0.061 & -0.045 & -0.038 \\ \midrule
\textbf{LAA} & ~ & ~ & ~ & ~ & ~ & ~ & ~ & ~ & ~ & ~ \\ \midrule
Opp TBR (home) & -0.007 & -0.002 & -0.024 & -0.014 & 0.030 & 0.003 & -0.017 & -0.046 & -0.016 & -0.018 \\ 
Opp TBR (away) & 0.022 & 0.017 & 0.045 & 0.012 & 0.040 & 0.017 & 0.017 & -0.042 & -0.013 & -0.045 \\ 
Team TBR (home) & -0.003 & -0.022 & 0.013 & -0.014 & 0.030 & 0.012 & -0.016 & -0.028 & -0.015 & -0.027 \\ 
Team TBR (away) & 0.004 & 0.004 & 0.006 & 0.003 & -0.008 & 0.015 & -0.014 & -0.051 & -0.007 & -0.072 \\ \midrule
\textbf{LAD} & ~ & ~ & ~ & ~ & ~ & ~ & ~ & ~ & ~ & ~ \\ \midrule
Opp TBR (home) & 0.022 & 0.012 & 0.026 & 0.021 & 0.033 & -0.027 & -0.027 & -0.008 & -0.038 & -0.041 \\ 
Opp TBR (away) & 0.040 & 0.031 & 0.066 & 0.022 & 0.024 & -0.020 & -0.026 & -0.068 & -0.022 & -0.041 \\ 
Team TBR (home) & 0.019 & 0.003 & 0.049 & 0.006 & 0.047 & 0.002 & 0.001 & -0.004 & -0.016 & -0.027 \\ 
Team TBR (away) & -0.008 & 0.000 & 0.010 & 0.015 & 0.042 & -0.024 & -0.039 & -0.052 & -0.026 & -0.029 \\ \midrule
\textbf{MIA} & ~ & ~ & ~ & ~ & ~ & ~ & ~ & ~ & ~ & ~ \\ \midrule
Opp TBR (home) & -0.021 & 0.004 & 0.016 & -0.029 & 0.032 & 0.032 & -0.038 & -0.017 & -0.025 & -0.022 \\ 
Opp TBR (away) & 0.005 & 0.004 & 0.035 & 0.033 & 0.024 & 0.006 & 0.002 & 0.004 & 0.018 & -0.040 \\ 
Team TBR (home) & 0.004 & -0.009 & 0.046 & -0.036 & -0.001 & -0.017 & -0.021 & -0.046 & -0.022 & -0.056 \\ 
Team TBR (away) & 0.020 & 0.033 & 0.046 & 0.002 & 0.002 & 0.028 & 0.006 & -0.061 & -0.047 & -0.019 \\ \midrule
\textbf{MIL} & ~ & ~ & ~ & ~ & ~ & ~ & ~ & ~ & ~ & ~ \\ \midrule
Opp TBR (home) & 0.052 & 0.038 & 0.061 & 0.004 & 0.038 & 0.049 & -0.017 & -0.015 & -0.042 & -0.054 \\ 
Opp TBR (away) & 0.025 & 0.036 & 0.038 & -0.015 & 0.034 & 0.015 & -0.032 & -0.018 & -0.029 & -0.063 \\ 
Team TBR (home) & 0.042 & 0.057 & 0.082 & 0.009 & 0.043 & -0.016 & -0.019 & -0.025 & 0.000 & 0.009 \\ 
Team TBR (away) & 0.016 & 0.014 & 0.060 & 0.017 & 0.013 & -0.027 & -0.008 & -0.035 & -0.045 & -0.017 \\ \midrule
\textbf{MIN} & ~ & ~ & ~ & ~ & ~ & ~ & ~ & ~ & ~ & ~ \\ \midrule
Opp TBR (home) & -0.022 & 0.014 & 0.034 & -0.024 & 0.003 & -0.068 & -0.056 & -0.057 & -0.008 & -0.017 \\ 
Opp TBR (away) & 0.014 & 0.028 & 0.012 & 0.019 & 0.022 & 0.013 & 0.013 & -0.020 & -0.012 & -0.039 \\ 
Team TBR (home) & 0.043 & 0.043 & 0.054 & 0.012 & 0.002 & 0.033 & -0.072 & -0.072 & -0.011 & -0.016 \\ 
Team TBR (away) & 0.010 & 0.033 & 0.029 & 0.009 & 0.048 & -0.011 & -0.051 & -0.074 & -0.041 & -0.050 \\ \midrule
\textbf{NYM} & ~ & ~ & ~ & ~ & ~ & ~ & ~ & ~ & ~ & ~ \\ \midrule
Opp TBR (home) & 0.021 & 0.008 & 0.054 & 0.011 & 0.066 & 0.054 & -0.037 & -0.044 & -0.040 & -0.048 \\ 
Opp TBR (away) & 0.029 & 0.008 & 0.068 & 0.034 & 0.034 & 0.011 & -0.015 & -0.022 & -0.021 & -0.052 \\ 
Team TBR (home) & 0.011 & 0.011 & 0.027 & -0.009 & 0.024 & 0.046 & -0.032 & -0.047 & -0.041 & -0.046 \\ 
Team TBR (away) & -0.004 & 0.031 & 0.073 & 0.034 & 0.033 & 0.037 & -0.024 & -0.018 & -0.044 & -0.043 \\ \midrule
\textbf{NYY} & ~ & ~ & ~ & ~ & ~ & ~ & ~ & ~ & ~ & ~ \\ \midrule
Opp TBR (home) & 0.020 & 0.050 & -0.008 & 0.026 & 0.055 & -0.012 & -0.005 & -0.036 & -0.034 & -0.035 \\ 
Opp TBR (away) & 0.055 & 0.046 & 0.040 & -0.001 & 0.061 & 0.061 & -0.024 & -0.040 & -0.026 & -0.036 \\ 
Team TBR (home) & 0.051 & 0.045 & 0.066 & 0.028 & 0.033 & 0.054 & -0.056 & -0.056 & -0.076 & -0.072 \\ 
Team TBR (away) & 0.003 & 0.013 & 0.016 & 0.005 & 0.068 & -0.040 & -0.050 & -0.050 & -0.065 & -0.059 \\ \midrule
\textbf{OAK} & ~ & ~ & ~ & ~ & ~ & ~ & ~ & ~ & ~ & ~ \\ \midrule
Opp TBR (home) & -0.046 & -0.023 & 0.011 & -0.058 & -0.031 & -0.059 & -0.058 & -0.046 & -0.018 & -0.053 \\ 
Opp TBR (away) & 0.024 & 0.032 & 0.029 & -0.017 & 0.008 & 0.045 & -0.024 & -0.020 & 0.011 & -0.003 \\ 
Team TBR (home) & 0.042 & -0.034 & 0.044 & -0.041 & 0.002 & -0.037 & -0.059 & -0.048 & -0.032 & -0.064 \\ 
Team TBR (away) & 0.027 & 0.008 & -0.001 & 0.021 & 0.017 & -0.002 & -0.036 & -0.044 & -0.024 & -0.033 \\ \midrule
\textbf{PHI} & ~ & ~ & ~ & ~ & ~ & ~ & ~ & ~ & ~ & ~ \\ \midrule
Opp TBR (home) & 0.027 & 0.037 & 0.059 & 0.046 & 0.037 & 0.056 & 0.021 & -0.015 & -0.023 & -0.011 \\ 
Opp TBR (away) & 0.029 & 0.039 & 0.042 & 0.027 & 0.031 & 0.076 & 0.029 & -0.015 & -0.004 & -0.034 \\ 
Team TBR (home) & 0.025 & 0.023 & 0.044 & 0.034 & 0.050 & -0.007 & 0.000 & -0.023 & 0.011 & -0.003 \\ 
Team TBR (away) & 0.017 & 0.025 & 0.044 & -0.004 & 0.015 & 0.004 & -0.011 & -0.057 & -0.024 & -0.034 \\ \midrule
\textbf{PIT} & ~ & ~ & ~ & ~ & ~ & ~ & ~ & ~ & ~ & ~ \\ \midrule
Opp TBR (home) & -0.026 & 0.030 & 0.039 & -0.010 & 0.072 & -0.041 & -0.020 & -0.062 & -0.058 & -0.055 \\ 
Opp TBR (away) & 0.012 & 0.050 & 0.046 & 0.005 & 0.058 & 0.050 & 0.022 & -0.016 & -0.007 & -0.021 \\ 
Team TBR (home) & -0.011 & 0.027 & 0.031 & -0.008 & 0.010 & -0.022 & -0.007 & -0.029 & -0.051 & -0.063 \\ 
Team TBR (away) & 0.011 & 0.010 & 0.024 & 0.003 & 0.034 & -0.044 & -0.033 & -0.020 & 0.002 & -0.049 \\ \midrule
\textbf{SD} & ~ & ~ & ~ & ~ & ~ & ~ & ~ & ~ & ~ & ~ \\ \midrule
Opp TBR (home) & 0.069 & 0.013 & 0.037 & 0.007 & 0.005 & -0.068 & -0.060 & -0.071 & -0.020 & -0.019 \\ 
Opp TBR (away) & 0.039 & 0.026 & 0.090 & 0.011 & 0.069 & -0.007 & 0.008 & -0.008 & -0.026 & -0.055 \\ 
Team TBR (home) & 0.026 & 0.038 & 0.067 & 0.015 & -0.002 & 0.003 & -0.013 & -0.042 & -0.042 & -0.050 \\ 
Team TBR (away) & 0.016 & 0.028 & 0.052 & 0.016 & 0.076 & -0.039 & -0.034 & -0.023 & -0.031 & -0.034 \\ \midrule
\textbf{SEA} & ~ & ~ & ~ & ~ & ~ & ~ & ~ & ~ & ~ & ~ \\ \midrule
Opp TBR (home) & -0.002 & 0.022 & 0.001 & -0.009 & 0.012 & -0.013 & -0.046 & -0.036 & -0.012 & -0.064 \\ 
Opp TBR (away) & 0.051 & 0.015 & 0.043 & -0.015 & 0.045 & 0.039 & -0.009 & -0.037 & -0.043 & -0.033 \\ 
Team TBR (home) & -0.005 & -0.004 & 0.029 & -0.016 & 0.015 & 0.011 & -0.044 & -0.031 & -0.039 & -0.088 \\ 
Team TBR (away) & -0.017 & 0.008 & 0.025 & -0.020 & 0.069 & -0.019 & -0.031 & -0.061 & -0.029 & -0.045 \\ \midrule
\textbf{SF} & ~ & ~ & ~ & ~ & ~ & ~ & ~ & ~ & ~ & ~ \\ \midrule
Opp TBR (home) & -0.052 & -0.020 & -0.025 & -0.012 & -0.028 & -0.015 & -0.063 & -0.022 & -0.038 & -0.050 \\ 
Opp TBR (away) & 0.041 & 0.018 & 0.076 & -0.015 & 0.021 & -0.006 & -0.011 & 0.005 & 0.007 & -0.021 \\ 
Team TBR (home) & -0.003 & -0.002 & -0.013 & -0.010 & -0.046 & 0.021 & -0.034 & -0.047 & -0.062 & -0.060 \\ 
Team TBR (away) & 0.020 & 0.008 & 0.038 & -0.002 & 0.020 & 0.020 & 0.010 & -0.035 & -0.045 & -0.046 \\ \midrule
\textbf{STL} & ~ & ~ & ~ & ~ & ~ & ~ & ~ & ~ & ~ & ~ \\ \midrule
Opp TBR (home) & -0.007 & 0.004 & 0.023 & -0.026 & -0.027 & -0.068 & -0.079 & -0.072 & -0.047 & -0.064 \\ 
Opp TBR (away) & 0.000 & 0.051 & 0.058 & -0.005 & 0.030 & -0.027 & -0.050 & -0.028 & 0.014 & -0.015 \\ 
Team TBR (home) & 0.000 & -0.014 & -0.005 & -0.072 & -0.019 & -0.013 & -0.064 & -0.038 & -0.054 & -0.050 \\ 
Team TBR (away) & 0.022 & 0.051 & 0.045 & 0.026 & 0.026 & -0.037 & -0.015 & 0.008 & -0.029 & -0.033 \\ \midrule
\textbf{TB} & ~ & ~ & ~ & ~ & ~ & ~ & ~ & ~ & ~ & ~ \\ \midrule
Opp TBR (home) & -0.026 & -0.004 & 0.005 & -0.036 & 0.020 & 0.028 & -0.048 & -0.042 & -0.021 & -0.063 \\ 
Opp TBR (away) & -0.012 & 0.037 & 0.041 & 0.001 & 0.041 & 0.006 & -0.044 & -0.040 & -0.045 & -0.043 \\ 
Team TBR (home) & 0.028 & 0.027 & 0.062 & 0.034 & 0.012 & 0.058 & 0.026 & -0.011 & 0.026 & -0.031 \\ 
Team TBR (away) & 0.027 & 0.036 & 0.037 & 0.015 & 0.009 & 0.025 & 0.015 & -0.024 & 0.009 & -0.043 \\ \midrule
\textbf{TEX} & ~ & ~ & ~ & ~ & ~ & ~ & ~ & ~ & ~ & ~ \\ \midrule
Opp TBR (home) & 0.014 & 0.025 & 0.032 & 0.035 & 0.044 & 0.018 & -0.024 & -0.036 & -0.006 & -0.066 \\ 
Opp TBR (away) & 0.019 & -0.005 & 0.021 & -0.003 & 0.025 & 0.012 & -0.009 & -0.028 & -0.044 & -0.035 \\ 
Team TBR (home) & 0.016 & 0.052 & 0.046 & 0.037 & 0.041 & -0.015 & -0.023 & -0.018 & 0.018 & -0.037 \\ 
Team TBR (away) & 0.016 & 0.008 & 0.022 & -0.018 & 0.003 & -0.032 & -0.023 & -0.047 & -0.039 & -0.049 \\ \midrule
\textbf{TOR} & ~ & ~ & ~ & ~ & ~ & ~ & ~ & ~ & ~ & ~ \\ \midrule
Opp TBR (home) & -0.022 & 0.004 & 0.049 & 0.038 & 0.034 & 0.011 & -0.014 & -0.041 & -0.031 & -0.045 \\ 
Opp TBR (away) & 0.035 & -0.008 & 0.063 & 0.040 & 0.028 & 0.048 & -0.016 & -0.058 & -0.034 & -0.058 \\ 
Team TBR (home) & 0.039 & 0.008 & 0.023 & 0.002 & 0.043 & 0.006 & 0.004 & -0.023 & -0.023 & -0.031 \\ 
Team TBR (away) & 0.036 & 0.011 & 0.011 & -0.026 & 0.019 & -0.009 & -0.033 & -0.032 & -0.023 & -0.054 \\ \midrule
\textbf{WSH} & ~ & ~ & ~ & ~ & ~ & ~ & ~ & ~ & ~ & ~ \\ \midrule
Opp TBR (home) & -0.013 & -0.008 & 0.045 & 0.022 & 0.019 & -0.001 & -0.011 & -0.028 & -0.066 & -0.052 \\ 
Opp TBR (away) & 0.031 & 0.033 & 0.028 & -0.005 & 0.002 & 0.045 & -0.011 & -0.009 & -0.019 & -0.026 \\ 
Team TBR (home) & -0.003 & -0.002 & 0.062 & 0.009 & 0.050 & -0.005 & -0.028 & -0.046 & -0.009 & -0.045 \\ 
Team TBR (away) & 0.001 & 0.012 & 0.038 & -0.007 & 0.004 & 0.040 & -0.012 & -0.021 & -0.019 & -0.010 \\ 
\bottomrule
\end{longtable}
\endgroup

\newpage
% --- Appendix E ---
{
\section{TBR-PF$^{*}$ with 95\% confidence intervals} \label{app:park_ci}
\setcounter{table}{0}
\renewcommand{\thetable}{E.\arabic{table}}
\setcounter{figure}{0}
\renewcommand{\thefigure}{E.\arabic{figure}}

Table~\ref{tab:park_ci} reports the TBR-PF$^{*}$ estimates from
Table~\ref{tab:std_comparison_side_by_side} together with their 95\%
confidence intervals. For a centered ballpark-effect estimate $\tilde{\beta}^{\mathrm{park}}_p$, the interval on the standardized scale is computed as
\[
100+20\,\frac{\tilde{\beta}^{\mathrm{park}}_p\pm 1.96\,\mathrm{SE}
\bigl(\tilde{\beta}^{\mathrm{park}}_p\bigr)}{s_{\beta,\mathrm{park}}},
\]
where $s_{\beta,\mathrm{park}}$ is the standard deviation of the estimated ballpark effects in the corresponding season.

Figure~\ref{fig:park_ci_halfwidth} summarizes the distribution of the
95\% confidence-interval half-widths across the 300 park-season estimates.
The average half-width is 19.97 index points (SD = 2.94).

\begin{table}[htbp]
\footnotesize\sf\centering
\caption{TBR-PF$^{*}$ estimates with 95\% confidence intervals, 2015--2024. Each entry reports the standardized index defined in equation~\eqref{eq:index} with its 95\% confidence interval in parentheses. Rows are ordered by the ten-year average, from most to least conducive to total bases.\label{tab:park_ci}}
\vspace{1ex}

% 稍微縮減行高與欄距，以爭取單頁的垂直容納空間
\setlength{\tabcolsep}{3pt}
\renewcommand{\arraystretch}{0.85} 

% ====== 上半部：2015 -- 2019 ======
\begin{adjustbox}{max width=\textwidth}
\begin{tabular}{lrrrrr}
\toprule
Team & 2015 & 2016 & 2017 & 2018 & 2019 \\
\midrule
COL & 150 (128.1, 171.0) & 152 (133.5, 170.0) & 165 (147.7, 182.7) & 158 (141.1, 174.8) & 151 (135.2, 167.6) \\
CIN & 122 (100.3, 143.9) & 127 (108.2, 145.8) & 134 (115.8, 151.4) & 141 (124.0, 157.7) & 145 (127.7, 162.0) \\
HOU & 154 (131.4, 176.9) & 107 (87.9, 126.0) & 116 (98.7, 133.7) & 119 (101.9, 136.5) & 140 (123.5, 156.9) \\
BOS & 119 (97.9, 140.9) & 136 (117.9, 154.5) & 97 (79.2, 114.0) & 87 (70.7, 104.3) & 90 (73.9, 106.9) \\
MIL & 112 (89.7, 134.1) & 116 (97.1, 135.7) & 123 (105.1, 141.4) & 109 (91.1, 125.9) & 103 (86.2, 120.4) \\
AZ  & 104 (82.3, 126.1) & 119 (100.7, 137.8) & 108 (89.8, 125.6) & 101 (83.9, 118.2) & 99 (82.0, 115.2) \\
CLE & 106 (84.1, 128.6) & 116 (97.2, 134.5) & 101 (83.2, 118.9) & 85 (67.8, 101.3) & 117 (99.9, 134.0) \\
TEX & 102 (80.3, 123.6) & 116 (97.6, 134.7) & 106 (88.8, 124.1) & 118 (101.6, 135.4) & 106 (89.7, 123.0) \\
PHI & 94 (71.7, 115.5) & 99 (80.1, 118.1) & 105 (87.1, 122.7) & 120 (103.0, 137.9) & 110 (93.0, 126.5) \\
TB  & 98 (75.6, 120.9) & 93 (73.8, 112.5) & 102 (84.3, 120.6) & 96 (78.7, 113.3) & 85 (68.1, 102.1) \\
CWS & 109 (86.9, 131.1) & 115 (96.7, 134.2) & 116 (98.5, 134.2) & 104 (87.3, 121.7) & 89 (71.6, 105.6) \\
LAD & 93 (71.0, 116.0) & 89 (70.2, 108.6) & 94 (75.7, 112.1) & 105 (87.6, 122.2) & 118 (101.1, 134.8) \\
TOR & 99 (77.4, 121.1) & 100 (81.6, 119.2) & 90 (72.5, 107.9) & 92 (74.6, 108.5) & 102 (85.2, 118.9) \\
CHC & 100 (76.7, 122.6) & 91 (71.8, 110.2) & 104 (86.5, 122.3) & 110 (92.6, 126.4) & 99 (81.7, 115.5) \\
BAL & 119 (97.2, 141.3) & 94 (74.8, 112.4) & 92 (74.1, 109.0) & 105 (87.9, 121.5) & 112 (95.3, 128.2) \\
SD  & 110 (87.9, 133.0) & 106 (86.6, 124.5) & 103 (84.9, 121.3) & 123 (105.7, 140.0) & 81 (63.4, 98.0) \\
NYY & 113 (90.6, 134.6) & 112 (93.3, 131.1) & 99 (80.9, 116.9) & 107 (90.1, 124.5) & 96 (79.3, 113.4) \\
MIN & 101 (79.7, 123.2) & 103 (84.4, 121.2) & 112 (95.1, 129.9) & 95 (78.3, 112.2) & 84 (68.1, 100.9) \\
LAA & 85 (63.3, 107.6) & 82 (64.1, 100.9) & 73 (55.2, 90.4) & 83 (66.3, 100.6) & 106 (89.8, 123.0) \\
NYM & 88 (65.8, 110.5) & 90 (71.4, 109.3) & 91 (73.9, 109.1) & 91 (73.7, 108.3) & 103 (86.7, 120.3) \\
WSH & 76 (53.7, 98.5) & 77 (57.9, 95.6) & 112 (94.5, 129.8) & 109 (92.0, 125.7) & 111 (94.8, 128.2) \\
PIT & 70 (48.4, 92.3) & 120 (101.2, 138.4) & 98 (80.9, 115.7) & 97 (80.5, 114.3) & 100 (83.5, 116.0) \\
ATL & 66 (43.9, 88.0) & 73 (54.5, 91.5) & 98 (80.9, 115.5) & 93 (75.7, 109.6) & 94 (77.2, 110.9) \\
MIA & 78 (56.4, 100.2) & 82 (62.7, 100.4) & 95 (77.3, 112.2) & 74 (57.0, 90.5) & 92 (75.6, 109.4) \\
SEA & 79 (56.7, 101.3) & 91 (72.5, 110.0) & 87 (69.5, 104.8) & 90 (73.0, 107.1) & 87 (69.5, 103.6) \\
KC  & 97 (75.5, 118.7) & 80 (61.2, 98.1) & 82 (64.9, 99.8) & 73 (56.7, 89.9) & 79 (62.8, 95.8) \\
SF  & 81 (58.8, 102.8) & 98 (80.0, 116.8) & 57 (40.2, 74.7) & 102 (84.9, 118.7) & 65 (48.4, 81.9) \\
DET & 79 (57.1, 100.4) & 74 (55.1, 92.6) & 70 (53.1, 87.6) & 72 (55.2, 88.7) & 79 (61.9, 95.4) \\
OAK & 95 (73.1, 116.3) & 67 (48.1, 85.2) & 95 (77.7, 113.2) & 71 (53.9, 87.8) & 88 (71.4, 104.5) \\
STL & 99 (76.8, 121.0) & 74 (55.2, 92.5) & 71 (53.5, 89.1) & 70 (53.0, 86.9) & 68 (50.6, 84.4) \\
\bottomrule
\end{tabular}
\end{adjustbox}

\vspace{1.0em} % 兩表之間的垂直間距

% ====== 下半部：2020 -- 2024 ======
\begin{adjustbox}{max width=\textwidth}
\begin{tabular}{lrrrrr}
\toprule
Team & 2020 & 2021 & 2022 & 2023 & 2024 \\
\midrule
COL & 143 (118.9, 167.7) & 144 (124.7, 162.4) & 151 (131.8, 170.5) & 135 (116.0, 154.8) & 146 (124.7, 166.8) \\
CIN & 132 (104.4, 158.9) & 139 (119.4, 158.3) & 151 (130.4, 171.2) & 145 (124.8, 164.6) & 149 (127.5, 170.9) \\
HOU & 92 (66.7, 117.7) & 107 (88.5, 126.5) & 114 (93.8, 134.2) & 100 (80.7, 119.8) & 115 (93.9, 136.1) \\
BOS & 153 (123.7, 182.5) & 105 (86.3, 124.5) & 121 (100.7, 140.5) & 120 (100.3, 139.1) & 119 (97.2, 140.2) \\
MIL & 97 (69.5, 124.0) & 105 (84.7, 124.6) & 104 (83.3, 124.5) & 107 (87.2, 127.7) & 126 (104.4, 148.2) \\
AZ  & 99 (74.3, 124.1) & 133 (114.3, 152.4) & 101 (80.6, 120.7) & 127 (107.6, 146.6) & 105 (84.3, 125.7) \\
CLE & 113 (86.9, 139.5) & 114 (93.9, 133.3) & 108 (88.2, 127.6) & 88 (68.5, 107.2) & 120 (98.2, 141.3) \\
TEX & 91 (65.6, 117.3) & 92 (73.2, 111.6) & 109 (89.0, 129.3) & 131 (110.9, 150.2) & 90 (68.6, 111.3) \\
PHI & 90 (61.7, 118.2) & 108 (88.6, 127.7) & 100 (80.0, 120.3) & 112 (92.3, 132.0) & 123 (101.7, 144.0) \\
TB  & 125 (93.7, 155.9) & 116 (96.6, 136.2) & 115 (94.8, 135.5) & 131 (111.0, 151.4) & 97 (75.4, 118.8) \\
CWS & 115 (88.5, 140.6) & 125 (105.0, 144.4) & 95 (75.4, 115.0) & 95 (75.5, 115.1) & 88 (66.8, 109.5) \\
LAD & 105 (79.5, 129.7) & 110 (90.5, 130.0) & 130 (110.2, 150.8) & 99 (79.6, 119.1) & 106 (84.6, 127.1) \\
TOR & 99 (69.0, 129.9) & 113 (93.5, 131.7) & 113 (93.0, 132.8) & 103 (83.5, 123.0) & 106 (84.4, 126.8) \\
CHC & 78 (51.4, 104.5) & 118 (97.8, 137.6) & 114 (94.0, 134.5) & 115 (95.1, 134.6) & 80 (58.4, 101.7) \\
BAL & 106 (76.4, 136.4) & 110 (90.9, 129.2) & 80 (60.3, 100.4) & 75 (55.3, 95.0) & 90 (69.1, 111.4) \\
SD  & 90 (63.6, 116.0) & 89 (69.8, 108.9) & 78 (57.8, 98.5) & 93 (72.9, 112.7) & 101 (80.2, 122.3) \\
NYY & 111 (80.9, 141.6) & 84 (64.4, 104.3) & 95 (74.5, 115.5) & 73 (52.9, 93.2) & 82 (60.7, 103.7) \\
MIN & 94 (67.5, 120.4) & 59 (40.0, 78.6) & 74 (53.5, 93.6) & 108 (87.2, 128.2) & 116 (95.0, 137.3) \\
LAA & 95 (70.3, 120.1) & 92 (72.8, 111.6) & 105 (84.6, 125.5) & 102 (81.6, 121.9) & 112 (90.2, 133.4) \\
NYM & 120 (91.8, 148.7) & 87 (66.9, 107.1) & 81 (60.8, 101.2) & 86 (65.9, 106.0) & 95 (73.2, 116.3) \\
WSH & 83 (54.4, 111.4) & 95 (75.9, 114.4) & 84 (63.7, 103.4) & 95 (75.5, 113.6) & 89 (67.5, 109.6) \\
PIT & 71 (45.7, 96.9) & 99 (79.8, 118.4) & 89 (69.3, 109.7) & 74 (54.1, 93.9) & 80 (58.5, 101.2) \\
ATL & 97 (82.6, 111.9) & 98 (78.0, 117.4) & 82 (61.5, 102.8) & 101 (81.5, 120.8) & 85 (63.1, 106.7) \\
MIA & 93 (63.5, 122.3) & 90 (69.9, 109.3) & 83 (62.5, 103.2) & 91 (71.7, 110.8) & 96 (75.3, 117.2) \\
SEA & 86 (60.2, 112.6) & 78 (58.2, 97.7) & 106 (85.4, 126.3) & 101 (81.0, 121.5) & 63 (40.6, 85.7) \\
KC  & 89 (63.3, 114.8) & 87 (67.8, 105.8) & 79 (58.9, 98.2) & 75 (55.9, 94.6) & 91 (70.5, 112.4) \\
SF  & 107 (81.9, 132.1) & 83 (63.5, 102.5) & 87 (67.0, 107.7) & 70 (50.2, 89.9) & 79 (57.5, 100.1) \\
DET & 86 (59.5, 113.1) & 76 (56.1, 95.1) & 79 (58.4, 98.7) & 88 (68.3, 107.9) & 96 (74.1, 117.1) \\
OAK & 60 (34.1, 85.9) & 71 (51.6, 90.6) & 88 (67.9, 108.2) & 89 (68.7, 109.0) & 72 (50.5, 93.2) \\
STL & 78 (50.8, 104.9) & 73 (53.2, 92.0) & 84 (64.5, 104.1) & 70 (50.8, 89.2) & 84 (62.8, 104.9) \\
\bottomrule
\end{tabular}
\end{adjustbox}

\end{table}

\begin{figure}[htbp]
    \centering
    \includegraphics[width=0.8\columnwidth]{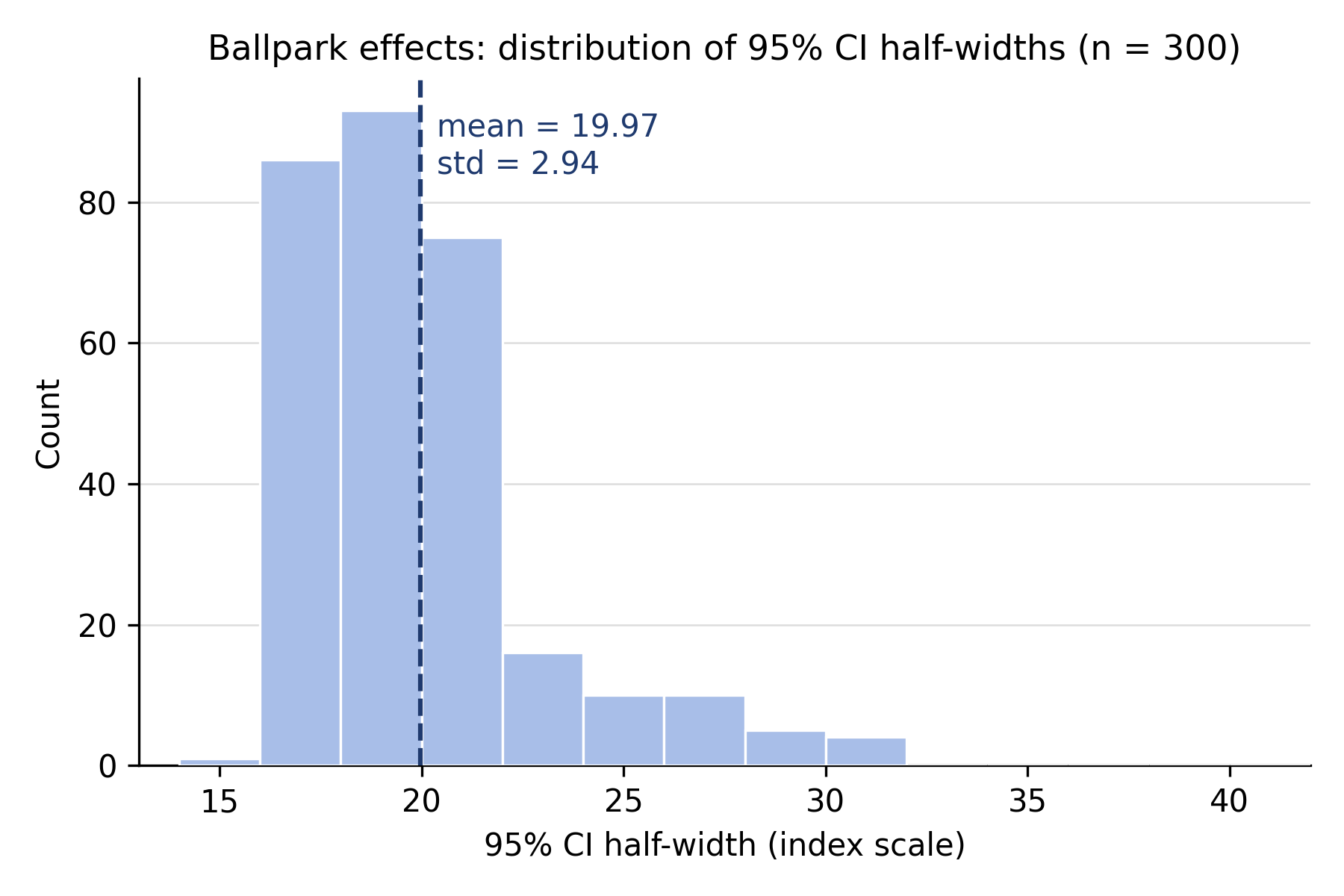}
    \caption{Distribution of the 95\% confidence-interval half-widths of TBR-PF$^{*}$ in Table~\ref{tab:park_ci}
    (30 ballparks $\times$ 10 seasons; mean $=19.97$, SD $=2.94$).}
    \label{fig:park_ci_halfwidth}
\end{figure}

\subsection*{A simple benchmark for the size of the confidence intervals}

The magnitude of these confidence intervals is consistent with the variability of individual batted-ball outcomes and the amount of information available in one season. The standard deviation of individual TBR values is very stable across seasons, ranging from 0.679 to 0.715, with an average of 0.696. In a full season, approximately 4,000 batted balls are observed in each ballpark. If a park effect were estimated by a simple park-specific average, a rough benchmark for its standard error would therefore be
\[
\frac{0.696}{\sqrt{4000}}
\approx 0.011.
\]

This value is close to the regression-based standard errors, which are generally about 0.013 to 0.015 in the full seasons. The slightly larger regression-based values are reasonable because the model estimates ballpark and defensive-team effects at the same time and because the schedule does not provide a perfectly balanced set of defensive teams in every ballpark.

The shortened 2020 season provides an additional check. In that season, each ballpark contains only about 1,500 batted-ball observations. Using the 2020 TBR standard deviation of approximately 0.702 gives
\[
\frac{0.702}{\sqrt{1500}} \approx 0.018,
\]
which is consistent with the larger regression-based standard errors observed
for 2020.

These calculations are intended only as a simple benchmark; the confidence intervals in Table~\ref{tab:park_ci} are based on the full weighted least-squares covariance matrix under the centered parameterization. Nevertheless, the close agreement in magnitude indicates that the reported uncertainty is consistent with the intrinsic variability of batted-ball outcomes and the finite sample available within a single season. The uncertainty estimates therefore provide a quantitative explanation for why annual park-factor estimates may fluctuate even when the physical characteristics of a ballpark remain largely unchanged.

\newpage
% --- Appendix F ---
\section{DBS$^{*}$ with 95\% confidence intervals} \label{app:defense_ci}
\setcounter{table}{0}
\renewcommand{\thetable}{F.\arabic{table}}
\setcounter{figure}{0}
\renewcommand{\thefigure}{F.\arabic{figure}}

Table~\ref{tab:defense_ci} reports the DBS$^{*}$ estimates from Table~\ref{tab:std_defense_comparison} together with their 95\% confidence intervals. For a centered team-defense estimate $\tilde{\beta}^{\mathrm{def}}_d$, the interval on the standardized scale is computed as
\[
100+20\,\frac{\tilde{\beta}^{\mathrm{def}}_d \pm 1.96\,\mathrm{SE}
\bigl(\tilde{\beta}^{\mathrm{def}}_d\bigr)}{s_{\beta,\mathrm{def}}},
\]
where $s_{\beta,\mathrm{def}}$ is the standard deviation of the estimated team-defense effects in the corresponding season.

Figure~\ref{fig:defense_ci_halfwidth} summarizes the distribution of the 95\% confidence-interval half-widths across the 300 team-season estimates. The average half-width is 30.84 index points (SD = 3.74).

\begin{figure}[htbp]
    \centering
    \includegraphics[width=0.8\columnwidth]{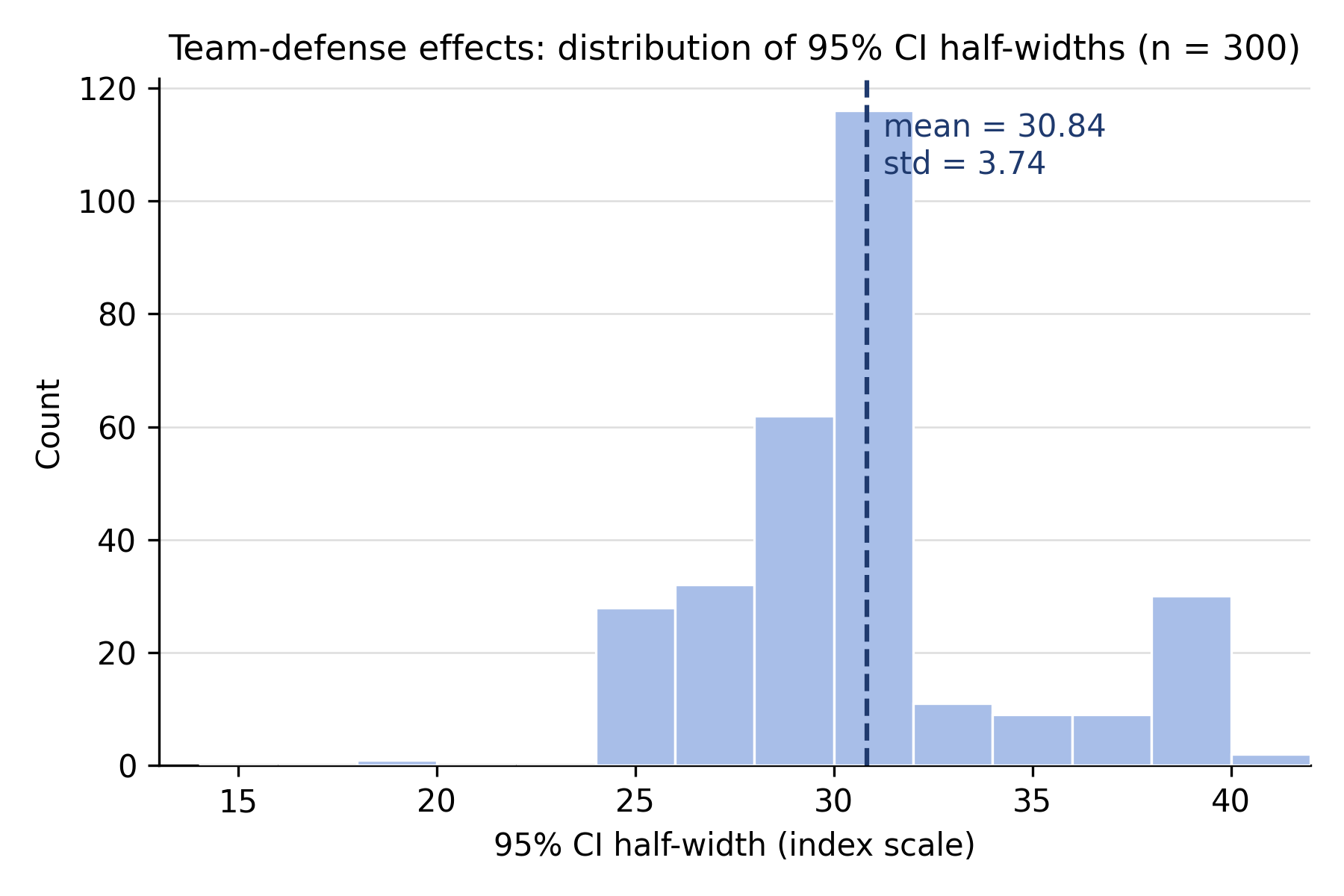}
    \caption{Distribution of the 95\% confidence-interval half-widths of DBS$^{*}$ in Table~\ref{tab:defense_ci}
    (30 teams $\times$ 10 seasons; mean $=30.84$, SD $=3.74$).}
    \label{fig:defense_ci_halfwidth}
\end{figure}

\begin{table}[htbp]
\footnotesize\sf\centering
\caption{DBS$^{*}$ estimates with 95\% confidence intervals, 2015--2024. Each entry reports the standardized index defined in equation~\eqref{eq:index} with its 95\% confidence interval in parentheses. Rows are ordered by the ten-year average, from strongest to weakest team defense.\label{tab:defense_ci}}
\vspace{1ex}

% 稍微縮減行高與欄距，以爭取單頁的垂直容納空間
\setlength{\tabcolsep}{3pt}
\renewcommand{\arraystretch}{0.85} 

% ====== 上半部：2015 -- 2019 ======
\begin{adjustbox}{max width=\textwidth}
\begin{tabular}{lrrrrr}
\toprule
Team & 2015 & 2016 & 2017 & 2018 & 2019 \\
\midrule
TB  & 143 (112.3, 173.3) & 109 (78.0, 139.8) & 114 (82.5, 145.2) & 115 (89.2, 140.6) & 85 (53.3, 116.6) \\
CLE & 121 (90.2, 152.2) & 111 (79.6, 141.4) & 121 (88.2, 153.1) & 60 (34.6, 85.9) & 115 (83.3, 146.4) \\
CHC & 98 (67.4, 129.4) & 136 (104.7, 167.7) & 127 (94.8, 158.8) & 135 (110.0, 160.7) & 123 (91.9, 154.2) \\
AZ  & 99 (69.5, 128.9) & 107 (76.9, 137.7) & 122 (89.8, 153.5) & 139 (113.3, 164.9) & 138 (107.3, 169.1) \\
LAD & 87 (56.4, 117.9) & 101 (69.9, 132.4) & 91 (58.1, 122.9) & 87 (60.9, 112.8) & 115 (83.5, 146.8) \\
TEX & 113 (83.3, 142.3) & 126 (96.0, 156.5) & 120 (89.5, 151.3) & 101 (76.2, 126.3) & 99 (68.1, 130.0) \\
COL & 79 (49.2, 108.3) & 115 (84.6, 145.0) & 112 (80.5, 142.7) & 113 (87.5, 138.7) & 126 (95.4, 156.0) \\
MIL & 87 (56.4, 116.9) & 96 (65.1, 126.0) & 111 (79.5, 142.9) & 120 (94.2, 145.8) & 93 (61.6, 123.5) \\
HOU & 132 (101.9, 162.9) & 72 (41.5, 102.6) & 78 (46.0, 109.7) & 103 (76.3, 129.5) & 113 (80.8, 144.5) \\
BOS & 101 (71.4, 130.5) & 150 (119.4, 181.2) & 117 (85.3, 148.5) & 89 (63.6, 115.2) & 80 (49.3, 111.5) \\
KC  & 122 (92.0, 151.5) & 113 (82.0, 143.2) & 88 (57.0, 118.3) & 82 (57.4, 106.9) & 104 (72.9, 134.2) \\
TOR & 117 (87.0, 146.2) & 129 (98.7, 160.0) & 64 (33.5, 95.2) & 59 (33.5, 83.7) & 102 (71.5, 133.4) \\
MIN & 133 (103.8, 162.4) & 99 (69.4, 129.3) & 126 (94.9, 156.2) & 87 (61.2, 111.9) & 101 (70.3, 131.8) \\
STL & 111 (80.3, 141.1) & 78 (48.0, 108.7) & 75 (44.0, 106.7) & 104 (78.7, 129.5) & 102 (70.3, 133.2) \\
SF  & 103 (72.7, 132.5) & 121 (90.4, 151.4) & 88 (57.3, 118.2) & 126 (100.4, 150.8) & 112 (80.8, 142.3) \\
SD  & 74 (43.4, 104.5) & 114 (83.9, 145.0) & 74 (42.8, 105.5) & 111 (85.2, 136.4) & 84 (52.3, 115.2) \\
OAK & 124 (94.8, 154.1) & 93 (62.8, 123.1) & 117 (85.9, 147.7) & 118 (92.4, 142.7) & 126 (95.1, 156.2) \\
WSH & 95 (64.4, 125.3) & 71 (40.0, 102.4) & 115 (83.2, 146.4) & 105 (79.6, 130.9) & 135 (103.9, 166.9) \\
ATL & 72 (42.2, 101.8) & 81 (50.6, 110.9) & 85 (54.1, 115.6) & 119 (92.8, 144.6) & 117 (86.5, 148.4) \\
BAL & 90 (59.6, 120.2) & 77 (46.1, 107.1) & 89 (58.1, 119.9) & 72 (47.5, 97.2) & 94 (64.1, 124.6) \\
LAA & 107 (76.8, 136.6) & 99 (68.7, 128.6) & 113 (82.1, 144.8) & 94 (68.5, 119.9) & 86 (54.9, 116.7) \\
CWS & 73 (42.7, 103.1) & 104 (73.7, 134.4) & 112 (80.9, 143.0) & 111 (85.8, 136.3) & 86 (55.2, 117.2) \\
SEA & 86 (56.2, 116.0) & 94 (63.7, 124.4) & 110 (79.1, 140.7) & 108 (82.7, 133.2) & 82 (51.3, 112.2) \\
CIN & 94 (64.6, 124.0) & 96 (65.3, 126.1) & 61 (30.3, 92.4) & 114 (88.7, 138.5) & 106 (73.7, 137.6) \\
NYY & 92 (62.3, 122.6) & 86 (54.8, 116.7) & 125 (93.5, 157.2) & 95 (69.2, 121.6) & 67 (36.1, 98.6) \\
MIA & 114 (84.4, 144.5) & 98 (67.5, 129.0) & 113 (82.7, 143.9) & 88 (62.6, 112.9) & 100 (69.2, 131.6) \\
DET & 62 (32.9, 91.7) & 73 (43.0, 103.8) & 81 (49.9, 111.3) & 72 (46.5, 96.8) & 67 (36.5, 98.4) \\
NYM & 80 (50.2, 110.7) & 101 (70.6, 131.8) & 67 (36.6, 98.1) & 79 (53.8, 104.6) & 78 (47.0, 108.5) \\
PIT & 110 (79.9, 139.8) & 73 (42.6, 102.8) & 94 (62.7, 124.7) & 109 (83.6, 134.1) & 56 (25.1, 86.5) \\
PHI & 80 (51.2, 109.8) & 77 (46.3, 107.3) & 91 (60.4, 122.3) & 86 (60.0, 111.1) & 108 (77.4, 139.3) \\
\bottomrule
\end{tabular}
\end{adjustbox}

\vspace{1.0em} % 兩表之間的垂直間距

% ====== 下半部：2020 -- 2024 ======
\begin{adjustbox}{max width=\textwidth}
\begin{tabular}{lrrrrr}
\toprule
Team & 2020 & 2021 & 2022 & 2023 & 2024 \\
\midrule
TB  & 112 (71.5, 152.2) & 139 (109.5, 167.9) & 118 (91.2, 145.4) & 132 (101.6, 162.2) & 121 (82.2, 159.4) \\
CLE & 115 (79.9, 150.3) & 124 (93.9, 153.3) & 144 (117.5, 171.5) & 127 (98.0, 156.6) & 117 (78.2, 156.5) \\
CHC & 94 (59.3, 128.1) & 91 (61.2, 120.4) & 99 (71.8, 125.6) & 128 (97.9, 157.7) & 108 (69.3, 145.8) \\
AZ  & 93 (60.0, 126.8) & 95 (66.8, 124.1) & 118 (91.0, 144.6) & 125 (94.9, 154.2) & 80 (42.1, 117.1) \\
LAD & 127 (93.8, 160.6) & 120 (90.0, 150.3) & 123 (95.7, 150.8) & 109 (79.7, 139.2) & 113 (74.7, 151.4) \\
TEX & 85 (52.1, 118.9) & 89 (60.5, 118.3) & 103 (75.8, 129.7) & 122 (92.4, 152.1) & 111 (72.7, 148.9) \\
COL & 121 (88.3, 153.0) & 102 (73.1, 131.5) & 109 (82.8, 135.7) & 91 (61.6, 119.9) & 101 (63.9, 138.2) \\
MIL & 70 (34.1, 105.3) & 112 (81.3, 142.6) & 86 (58.3, 113.5) & 120 (90.3, 150.7) & 157 (119.1, 195.9) \\
HOU & 89 (55.5, 122.2) & 116 (87.0, 145.9) & 133 (105.6, 160.8) & 119 (89.3, 149.5) & 96 (56.6, 134.6) \\
BOS & 121 (81.9, 160.4) & 73 (43.7, 102.9) & 115 (88.0, 142.0) & 79 (49.7, 109.3) & 121 (82.7, 159.3) \\
KC  & 112 (77.6, 146.5) & 134 (104.7, 163.4) & 90 (63.1, 116.0) & 69 (39.8, 99.1) & 125 (87.0, 163.8) \\
TOR & 92 (52.0, 131.5) & 102 (72.0, 131.8) & 124 (97.5, 151.5) & 112 (82.1, 142.2) & 123 (84.5, 161.3) \\
MIN & 120 (85.4, 154.9) & 73 (43.7, 102.2) & 85 (57.9, 111.6) & 89 (58.3, 119.6) & 97 (58.9, 136.0) \\
STL & 124 (87.7, 161.0) & 134 (104.8, 163.3) & 111 (84.7, 137.9) & 72 (42.6, 100.5) & 89 (50.5, 126.5) \\
SF  & 116 (82.7, 149.7) & 112 (82.7, 141.5) & 68 (40.5, 94.6) & 70 (40.2, 99.6) & 78 (39.0, 116.8) \\
SD  & 130 (95.7, 164.2) & 106 (76.6, 136.0) & 95 (67.4, 122.0) & 94 (64.2, 124.4) & 104 (65.9, 142.9) \\
OAK & 84 (49.7, 117.9) & 101 (72.0, 130.2) & 92 (65.5, 119.1) & 67 (37.3, 96.7) & 62 (24.4, 100.3) \\
WSH & 81 (43.6, 117.8) & 90 (60.9, 119.5) & 71 (44.1, 97.2) & 118 (89.1, 147.3) & 93 (55.4, 131.2) \\
ATL & 101 (81.8, 120.3) & 125 (94.8, 154.4) & 94 (66.3, 121.8) & 109 (79.2, 139.4) & 71 (31.6, 110.8) \\
BAL & 116 (76.0, 155.7) & 100 (71.1, 128.4) & 100 (73.5, 127.2) & 119 (88.7, 148.7) & 109 (70.9, 147.9) \\
LAA & 89 (55.7, 122.9) & 68 (38.8, 97.6) & 114 (86.8, 141.6) & 86 (56.2, 116.2) & 98 (60.0, 136.6) \\
CWS & 136 (101.8, 170.0) & 105 (74.9, 135.4) & 81 (54.3, 108.7) & 79 (49.1, 109.2) & 64 (25.6, 102.5) \\
SEA & 80 (46.3, 113.9) & 91 (62.3, 120.5) & 105 (77.9, 132.5) & 107 (77.2, 137.1) & 85 (45.3, 123.7) \\
CIN & 88 (52.1, 124.7) & 77 (47.1, 106.4) & 81 (53.1, 108.0) & 90 (60.2, 119.6) & 110 (71.5, 149.3) \\
NYY & 100 (59.0, 140.9) & 78 (48.0, 107.8) & 98 (71.0, 126.0) & 88 (57.4, 118.1) & 82 (43.7, 121.1) \\
MIA & 88 (50.6, 125.0) & 94 (64.9, 123.9) & 57 (29.6, 84.8) & 71 (41.4, 101.5) & 87 (48.9, 124.9) \\
DET & 96 (61.5, 129.5) & 107 (77.6, 135.6) & 124 (96.6, 150.5) & 119 (89.1, 149.0) & 97 (58.3, 135.4) \\
NYM & 92 (53.8, 130.0) & 100 (69.8, 130.0) & 84 (56.0, 111.1) & 96 (66.0, 126.1) & 114 (75.2, 153.2) \\
PIT & 82 (47.8, 116.8) & 82 (52.2, 111.0) & 100 (72.9, 126.7) & 94 (63.6, 123.6) & 86 (47.4, 124.4) \\
PHI & 46 (7.5, 84.1) & 59 (29.3, 88.3) & 78 (50.4, 105.0) & 97 (67.1, 126.7) & 100 (61.1, 137.9) \\
\bottomrule
\end{tabular}
\end{adjustbox}

\end{table}
}

\end{document}